\documentclass{aa} 

\usepackage{graphicx}
\usepackage[varg]{txfonts}
\usepackage{rotating}

\begin{document}

\title{A search for photometric variability in magnetic chemically peculiar stars using ASAS-3 data}

\author{K. Bernhard\inst{1,2}
\and S.~H{\"u}mmerich\inst{1,2}
\and S.~Otero\inst{2,3}
\and E.~Paunzen\inst{4}}
\institute{Bundesdeutsche Arbeitsgemeinschaft für Ver{\"a}nderliche Sterne e.V. (BAV), Berlin, Germany \\
\email{klaus.bernhard@liwest.at,ernham@rz-online.de}
\and American Association of Variable Star Observers (AAVSO), Cambridge, USA
\and Buenos Aires, Argentina
\and Department of Theoretical Physics and Astrophysics, Masaryk University,
Kotl\'a\v{r}sk\'a 2, 611\,37 Brno, Czech Republic}

\date{} 
\abstract
{The (magnetic) chemically peculiar (CP) stars of the upper main sequence are well-suited laboratories for investigating 
the influence of magnetic fields on the stellar surface because they produce abundance inhomogeneities (spots), which 
results in photometric variability that is explained in terms of the oblique rotator model. CP stars exhibiting this phenomenon 
are normally classified as $\alpha^{2}$ Canum Venaticorum (ACV) variables. It is important to increase the sample of 
known rotational periods among CP stars by discovering new ACV variables. An increased sample size will contribute to the 
understanding of the CP stars' evolution in time.}
{We aim at discovering new ACV variables in the public data of the third phase of the All Sky Automated Survey (ASAS-3). 
Furthermore, by analysis of the available photometric data, we intend to derive rotational periods of the stars.} 
{The ASAS-3 data were cross-correlated with the Catalogue of Ap, HgMn, and Am stars in order to analyse the light curves of 
bona fide CP and related stars. The light curves were downloaded and cleaned of outliers and data points with a flag indicating bad quality. Promising candidates
showing a larger scatter than observed for constant stars in the corresponding magnitude range
were searched for periodic signals using a standard Fourier technique. Objects exhibiting periodic signals 
well above the noise level were considered and visually inspected, whereas borderline cases were rejected.} 
{In total, we found 323 variables, from which 246 are reported here for the first time, and 77 were probably wrongly 
classified before. The observed variability pattern of most stars is in accordance with an ACV classification. For some 
cases, it is difficult to distinguish between the light curves of double-waved ACVs and the variability induced by orbital 
motion (ellipsoidal variables/eclipsing variables), especially for objects exhibiting very small amplitudes and/or significant 
scatter in their light curves. Thus, some eclipsing or rotating ellipsoidal variables might be present. However, we are confident 
that the given periods are the correct ones. There seems to be a possible weak correlation between 
the rotational period and colour, in the sense that cooler magnetic CP stars rotate more slowly. However, this 
correlation seems to disappear when correcting for the interstellar reddening.}  
{The next steps have to include a compilation of all available rotational periods from the literature and a detailed investigation of the
astrophysical parameters of these stars. This includes a determination of the individual masses, luminosities, ages, and inclination 
angles. However, this information cannot be straightforwardly determined from photometric data alone.}

\maketitle

\titlerunning{A Search for Photometric Variability in mCP Stars using ASAS-3 data}
\authorrunning{Bernhard et al.}

\section{Introduction}

Chemically peculiar (CP) stars are upper main sequence objects (spectral types early B to early F) 
whose spectra are characterized by abnormally strong (or weak) absorption lines that indicate peculiar 
surface elemental abundances. The observed chemical peculiarities are thought to arise from the diffusion 
of chemical elements due to the competition between radiative pressure and gravitational settling 
\citep{Rich00, Turc03}. CP stars constitute about 10\% of all upper 
main sequence stars and are commonly subdivided into four classes \citep{Pres74}: metallic line (or Am) stars (CP1), 
magnetic Ap stars (CP2), HgMn stars (CP3), and He-weak stars (CP4).

The CP2 or Ap stars, which are the subject of the present investigation, are notorious for exhibiting 
strong, globally organized magnetic fields of up to several tens of kiloGauss \citep{Koch11}. Their atmospheres are 
enriched in elements such as Si, Cr, Sr, or Eu and usually present surface abundance patches or spots 
\citep{Mich81, Koch10, Krti13}, leading to photometric variability, which is considered to be caused by 
rotational modulation and explained in terms of the oblique rotator model \citep{Stib50}. 

As a result, the 
observed photometric period is the rotational period of the star. Generally, CP1 stars are not expected 
to show rotational light variability; in relation to CP3 stars, this is still a matter of debate 
\citep{Paun13, More14}. Interestingly, in a recent study using ultra-precise Kepler photometry, 
\citet{Balo15} have found that most of the investigated Am stars exhibit light curves indicative of 
rotational modulation due to star spots. With an amplitude of up to $\sim$200 ppm, this kind of variability 
is reserved for high-precision (space) photometry, though.
Photometrically variable CP2 stars are also designated as $\alpha^{2}$ Canum Venaticorum (ACV) variables in the 
General Catalogue of Variable Stars \citep[][GCVS]{Samu07}. Some Ap stars also exhibit photometric variability in the period range 
of $\sim$5 -- 20 minutes (high-overtone, low-degree, and
non-radial pulsation modes). These stars are known as rapidly oscillating Ap (roAp) stars \citep{Kurt90} and are 
not the subject of the present paper.

Considerable effort has been devoted to the study of the photometric variability of the magnetic Ap stars by 
photometric investigations or by mining available sky survey data \citep{Manf86, Paun98, Paun11, Wrai12}. The present paper 
investigates the photometric variability of 
Ap stars using the publicly available observations from the third phase of the All Sky Automated Survey \citep[][ASAS-3]{Pojm02} 
and aims at finding previously unidentified ACV variables. Observations, target selection, data 
analysis, and interpretation are described in Sect. \ref{obs_ana}. Results are presented and discussed in Sect. \ref{results}, and we 
conclude in Sect. \ref{conclusions}.

{\renewcommand{\arraystretch}{1.2}
\begin{table}
\caption{Statistical information on the composition of the present sample.}
\label{table_statistics}
\begin{center}
\begin{tabular}{lr}
\hline
\hline 
Type & Number of objects \\
\hline
Newly discovered ACV variables  & 239 \\
Variables of undetermined type$^{1}$  &  61 \\
Misclassified variables$^{1}$ & 16 \\
EB/ELL star candidates & 7 \\
Whole sample & 323 \\
\hline
\multicolumn{2}{l}{$^{1}$ ... identified as ACV variables for the first time}
\end{tabular}
\end{center}
\end{table}
}

\begin{figure}
\begin{center}
\includegraphics[width=0.47\textwidth,natwidth=2360,natheight=3387]{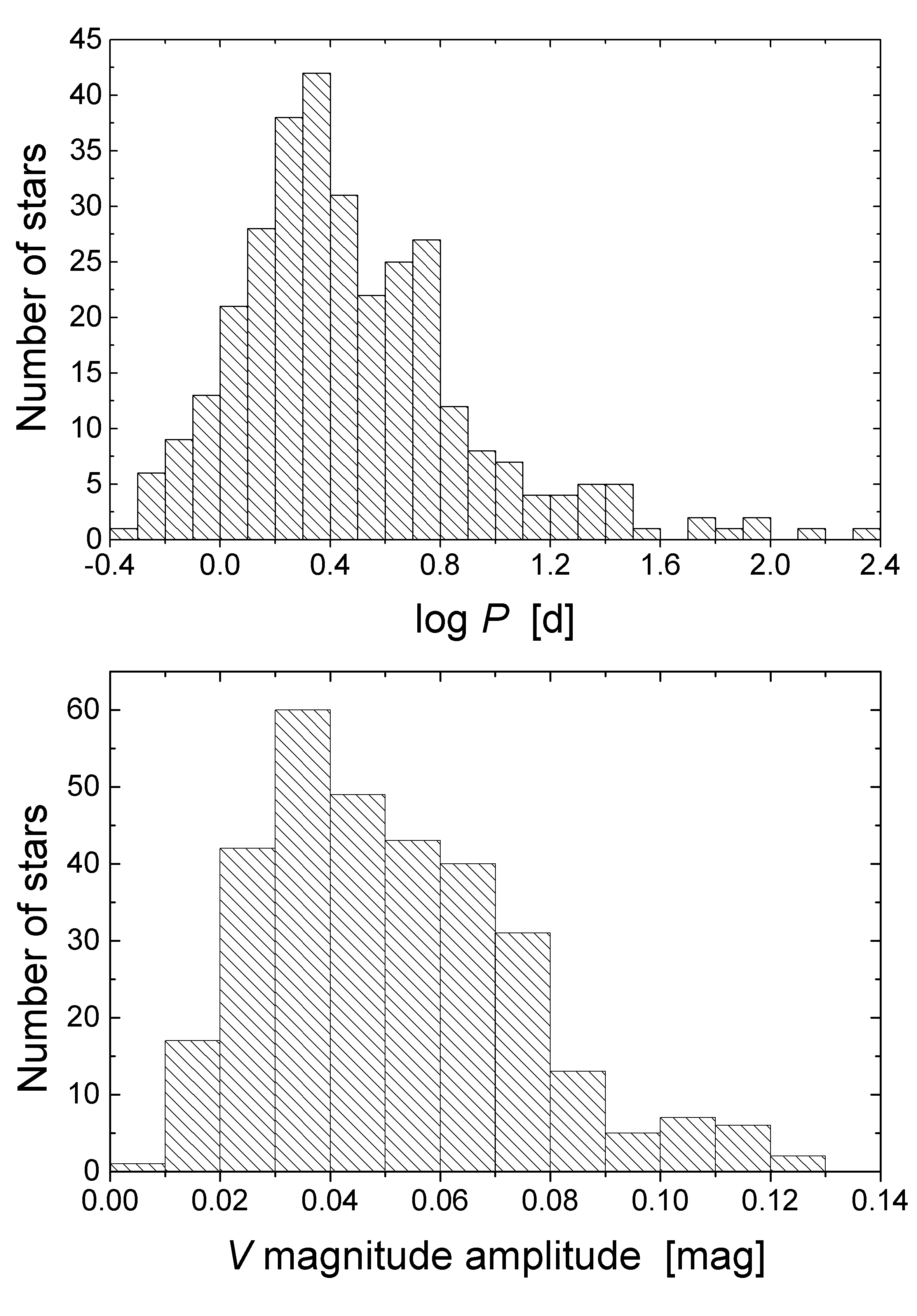}
\caption{Distribution of the logarithmic rotational periods (upper panel) and $V$ magnitude amplitudes (lower panel) among the photometrically variable Ap stars and Ap star candidates of the present sample (Table \ref{table_2}).} 
\label{distributions}
\end{center}
\end{figure}

\begin{figure*}
\begin{center}
\includegraphics[width=0.98\textwidth,natwidth=3390,natheight=2379]{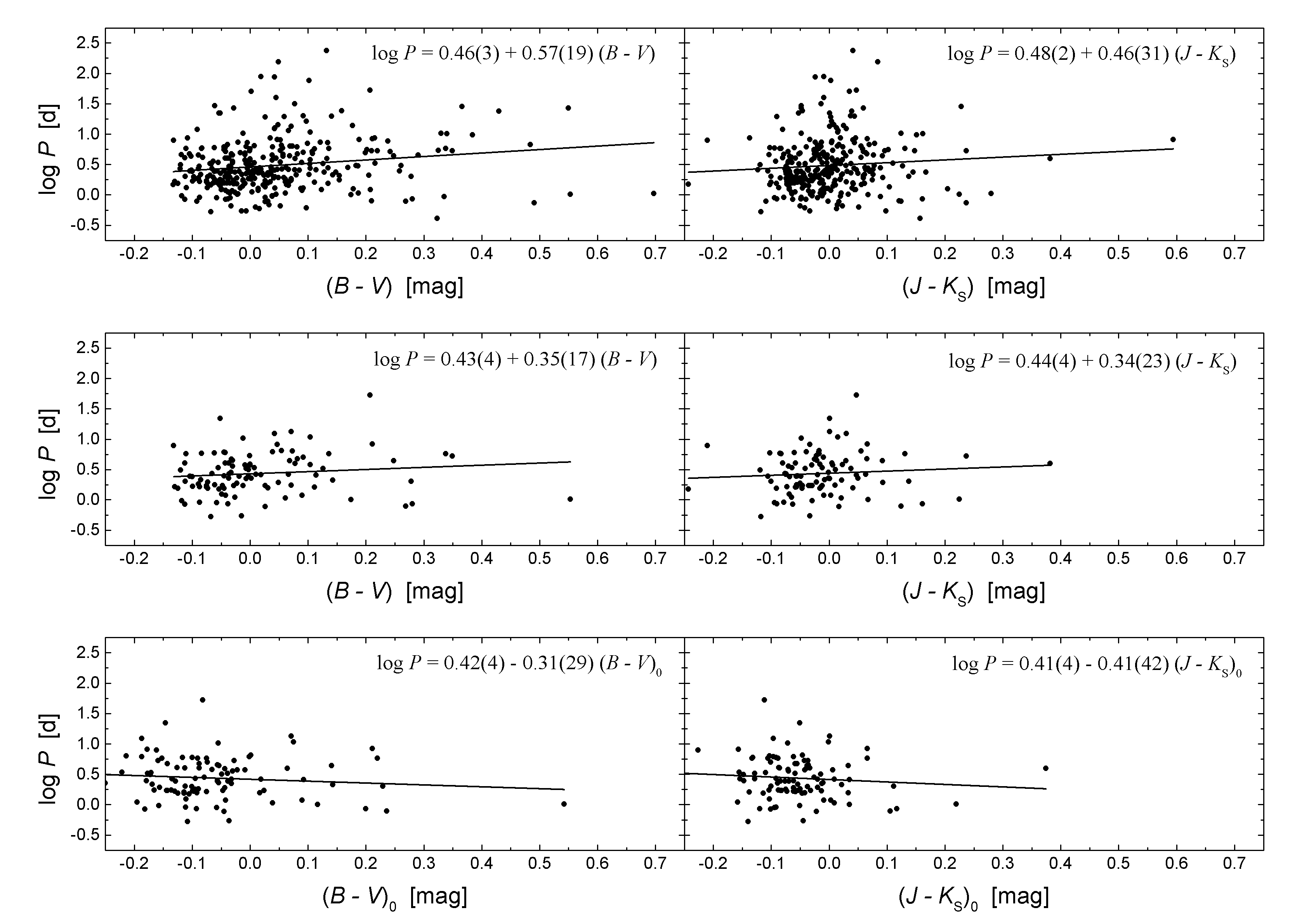}
\caption{Rotational periods as a function of $(B-V)$ and $(J-K_{\mathrm s})$ of the photometrically variable Ap stars and Ap star candidates of the complete sample (upper panels) and the 99 stars of our sample for which we were able to determine the reddening (middle and lower panels; Table \ref{reddening}).}
\label{bv_jk}
\end{center}
\end{figure*}

\section{Observations and analysis} \label{obs_ana}

The targets for the present investigation were primarily chosen among the Ap stars listed in
the most recent version of the Catalogue of Ap, HgMn, and Am stars \citep[][RM09 hereafter]{Rens09}. 
At the beginning of the project, a number of Am stars from the aforementioned list were inspected, too. 
According to our expectations, most of them did not exhibit variability in ASAS-3 data, which led us to abandon 
this approach and concentrate only on the Ap stars. The three Am stars that exhibit obvious light 
variability might be spectroscopically misclassified as Ap stars (Sect. \ref{results}).

The RM09 catalogue was cross-matched with the Tycho-2 catalogue \citep{Hog00}, and all objects with $V_{\mathrm T}$\,$<$\,11\,mag that 
have a reasonable number of observations in the ASAS-3 database were investigated. These are mostly southern objects with $\delta$\,$<$\,0\degr.
Additionally, other variables of spectral type late B to early A (with or without known chemical peculiarities) that have been investigated or 
discovered as by-products of other projects \citep{Oter07} and which exhibit photometric variability typical of ACVs were included in our investigation.

The GCVS, the AAVSO International Variable Star Index, VSX \citep{Wats06}, the VizieR \citep{Ochs00}, and SIMBAD \citep{Weng00} databases
were consulted to check for an entry in variability catalogues. Objects that have already been announced as ACV variables in the 
literature were dropped from our sample.

\subsection{Characteristics of the ASAS-3 data} \label{ASAS3}

The All Sky Automated Survey (ASAS) is a project that aims at continuous photometric monitoring of the whole sky, with the ultimate 
goal of detecting and investigating any kind of photometric variability. The first two phases of the project, 
ASAS-1 and ASAS-2, resulted in the discovery of about 3\,800 variable stars \citep{Pojm98, Pojm00}.

The third phase of the project, ASAS-3, which has produced a catalogue of about 50\,000 variable stars \citep{Pojm05}, 
lasted from 2000 until 2009 \citep{Pojm02} and has been monitoring the entire southern sky and part of the northern sky 
($\delta$\,$<$\,+28\degr). The ASAS-3 system, which occupied the ten-inch astrograph dome of the Las Campanas Observatory (Chile), 
consisted of two wide-field telescopes equipped with f/2.8 200\,mm Minolta lenses and 2048x2048 AP 10 Apogee detectors, 
covering a field of 8\fdg8x8\fdg8 of sky. Data were obtained through standard Johnson $I$ and $V$ filters during the ASAS-2 and ASAS-3 surveys, respectively. About 10$^{7}$ sources 
brighter than about $V$\,=\,14\,mag were catalogued. With a CCD resolution of about 14\farcs8 per pixel, the astrometric 
accuracy is around 3 -- 5\arcsec\ for bright stars and up to 15\arcsec\ for fainter stars. Thus, photometry in crowded fields as in
star clusters is rather uncertain.

The typical exposure time for ASAS-3 $V$-filter observations was three minutes, which resulted in reasonable photometry for stars in 
the magnitude range 7\,$\la$\,$V$\,$\la$\,14. The all-sky character of the survey necessitated sparse sampling. In general, a field was 
observed each one, two, or three days \citep{Pigu14} resulting in strong daily aliases in the corresponding Fourier spectra.

The most accurate photometry was obtained for stars in the magnitude range 8\,$\la$\,$V$\,$\la$\,10, exhibiting a typical scatter of 
about 0.01\,mag \citep{Pigu14}. However, because of the long time baseline of almost ten years, the detection limit of the 
ASAS-3 data for periodic signals is much lower than this value. \citet{Davi14} have detected periodic variables with amplitudes 
of variability of 0.01 -- 0.02\,mag in the magnitude range of 7\,$\la$\,$V$\,$\la$\,10. \citet{Pigu14} estimated that periodic signals with 
amplitudes as low as 4 - 5\,mmag can still be detected well. Thus, the ASAS-3 data are perfectly suited to an investigation of 
the low-amplitude photometric variability of CP stars \citep{Manf86}.

\begin{table*}
\caption{Essential data for the 316 stars identified as photometrically variable chemically peculiar stars or star candidates in the present paper.
Only part of the table is printed here for guidance regarding its form and content. The complete table is given in the appendix.}
\label{table_2}
\scriptsize{
\begin{center}
\begin{tabular}{llcclcccccc}
\hline
\hline 
Star &  ID (RM09)       & $\alpha$ (J2000) & $\delta$ (J2000)   & Type & Range ($V$) & Period & Epoch (HJD)      &  Spectral type &      $(B-V)$ & $(J-K_{\mathrm s})$ \\
     &              &                  &                  &      & [mag]       & [d]    & [d]         &               & [mag]   & [mag] \\
\hline
HD 10081        &       2510    &       01 36 15.100    &       $-$68 15 05.38   &       ACV     &       9.60$-$9.64     &       1.57032(3)      &       2452190.56(3)   &       A0 Sr Eu   &       +0.031  &        +0.041 \\
HD 26726        &       6830    &       04 12 46.635    &       $-$24 47 48.55   &       ACV     &       9.81$-$9.90     &       5.3816(4)       &       2451902.6(1)    &       A2 Sr      &       +0.325  &        +0.006 \\
HD 30898        &       7983    &       04 51 28.382    &       $-$00 32 47.85   &       ACV:    &       7.84$-$7.88     &       0.781776(8)     &       2452185.79(2)   &       A1$-$F0 &       +0.269  &        +0.124  \\
HD 35177        &       8980    &       05 23 01.934    &       +01 41 48.94    &       ACV     &       8.12$-$8.15     &       0.528194(5)     &       2452670.60(1)   &       B9 Si      &       $-$0.068        &       $-$0.118        \\
HD 36997        &       9810    &       05 35 13.819    &       $-$02 22 52.32   &       ACV     &       8.31$-$8.36     &       6.0072(4)       &       2453101.5(1)    &       B9 Si Sr   &       $-$0.043        &       $-$0.060        \\
HD 37189        &       9940    &       05 36 09.774    &       $-$13 49 20.35   &       ACV     &       8.53$-$8.57     &       2.8286(1)       &       2452115.90(6)   &       A0 Si      &       $-$0.033        &       $-$0.001        \\
HD 37713        &       10170   &       05 39 34.257    &       $-$25 08 53.44   &       ACV     &       8.11$-$8.14     &       1.50125(3)      &       2451914.64(3)   &       B9 Si      &       $-$0.054        &       $-$0.243        \\
HD 38366        &       10320   &       05 45 32.682    &       +09 19 00.77    &       ACV     &       8.94$-$9.01     &       0.625264(5)     &       2454143.68(1)   &       B9 Si      &       +0.009  &        +0.051 \\
HD 38698        &       10400   &       05 47 02.886    &       $-$14 28 26.75   &       ACV     &       9.09$-$9.12     &       3.1643(2)       &       2452690.64(6)   &       A0 Si      &       $-$0.003        &        +0.011 \\
HD 39635        &               &       05 53 25.382    &       $-$10 40 42.39   &       ACV:    &       8.75$-$8.82     &       3.9660(2)       &       2452752.51(8)   &       A0(V)   &       +0.078  &        +0.382  \\
HD 42382        &       11300   &       06 10 30.681    &       +03 28 58.72    &       ACV     &       9.11$-$9.15     &       5.9277(7)       &       2452921.9(2)    &       B9 Cr Eu Si        &       +0.056  &       $-$0.019        \\
HD 42695        &       11420   &       06 11 12.958    &       $-$22 49 41.46   &       ACV     &       8.12$-$8.14     &       1.06877(2)      &       2455148.78(3)   &       B9 Si      &       $-$0.035        &       $-$0.015        \\
HD 44038        &       11668   &       06 19 10.142    &       $-$09 22 06.21   &       ACV     &       9.73$-$9.77     &       0.689929(6)     &       2453029.75(1)   &       A0 Eu Cr Sr        &       +0.033  &        +0.014 \\
HD 44290        &       11750   &       06 20 28.159    &       $-$14 11 48.34   &       ACV     &       8.52$-$8.55     &       1.71523(4)      &       2452032.48(3)   &       B9 Cr Eu   &       +0.026  &       $-$0.031        \\
HD 44456        &       11800   &       06 21 55.420    &       +00 18 15.18    &       ACV     &       8.50$-$8.54     &       3.8934(2)       &       2452243.73(8)   &       A0 Si Sr   &       +0.000  &       $-$0.016        \\
HD 291674       &       12200   &       06 29 14.344    &       +00 42 57.01    &       ACV     &       9.75$-$9.78     &       2.7305(2)       &       2453457.57(8)   &       B9 Si      &       $-$0.005        &        +0.067 \\
HD 258583       &       12260   &       06 30 22.290    &       +08 39 36.98    &       ACV     &       8.76$-$8.83     &       10.312(1)       &       2452671.6(2)    &       A0 Si      &       $-$0.012        &       $-$0.048        \\
HD 45931        &       12280   &       06 30 27.489    &       $-$00 22 21.41   &       ACV     &       8.95$-$9.00     &       1.92699(5)      &       2452199.82(4)   &       A0 Si      &       +0.058  &       $-$0.017        \\
HD 46234        &       12370   &       06 30 32.021    &       $-$38 45 10.22   &       ACV     &       9.28$-$9.31     &       4.6936(4)       &       2451924.6(1)    &       A0 Si      &       $-$0.105        &       $-$0.052        \\
HD 47026        &       12590   &       06 33 58.822    &       $-$48 09 02.09   &       ACV     &       9.51$-$9.56     &       1.28037(2)      &       2452143.88(2)   &       B9 Sr Cr Eu        &       +0.014  &       $-$0.006        \\
\hline
\end{tabular}
\end{center}
}
\end{table*}

\subsection{Data analysis and interpretation} \label{data_int}

The light curves of all target stars were downloaded from the ASAS-3 website\footnote{http://www.astrouw.edu.pl/asas/}. No data from our targets exist in the ASAS-2 $I$-band catalogue. 
The light curves were inspected visually, and obvious outliers and data points with a quality flag of `D' (= `worst data, probably useless') were removed.
Promising candidates, i.e. stars showing a larger scatter than usually observed for apparent constant stars in the corresponding magnitude range with 
comparable instruments \citep{Hart04}, were searched for periodic signals in the frequency domain of 1\,$<$\,f (c/d)\,$<$\,30 using Period04 \citep{Lenz05}.

Objects exhibiting periodic signals well above the noise level (corresponding to a semi-amplitude of at least $\sim$0.007\,mag, as determined with Period04) 
were investigated. The data was folded with the resulting best-fitting frequency, and the light curve was visually inspected.
ASAS-3 data sometimes suffer from systematic trends. For example, strong blending effects may induce significant additional scatter due 
to the inclusion of part of the neighbouring star's flux, thus effectively rendering small-amplitude variations undetectable \citep{Site14}. 
Furthermore, instrumental long-term mean brightness trends might result in the detection of spurious periods. Consequently, objects whose light curves 
are indicative of strong systematic trends were rejected. Stars exhibiting convincing phase plots were kept.
Borderline cases -- that is to say, stars exhibiting a weak signal (semi-amplitude of $\le$\,0.007\,mag) 
that could not be attributed to systematic trends but did not produce a convincing phase plot either -- were rejected as well in order to keep the sample free of possibly 
spurious detections that might contaminate the sample of derived rotational periods.
The object was finally classified according to spectral type, colour information,
period, and shape of its light curve.
          
The observed variability pattern of most stars is in accordance with an ACV classification. Judging from light curve characteristics, some 
eclipsing or rotating ellipsoidal variables might be present, too. Because ACV variables are prone to exhibiting double-wave variations in their 
photometric light curves \citep{Mait80}, it is sometimes difficult to distinguish between the light curves of double-waved ACVs and 
the variability induced by orbital motion (ellipsoidal variables/eclipsing variables), especially for objects exhibiting very small amplitudes 
and/or significant scatter in their light curves.

Generally, there seems to be a lack of binaries among most subgroups of Ap stars \citep[for example, the Si, Si-Cr, and Si-Sr stars,][]{Gerb85, Leon99};
in particular, very few double-lined spectroscopic binaries comprising an Ap star are known \citep{Hubr14}, and a conspicuous lack of short-period 
systems has been described in the literature \citep{Nort04, Hubr05}.
Furthermore, the search for magnetic Ap stars in eclipsing binary systems has mostly been to no avail. Up to 2005, no such system 
has ever been found \citep{Hubr05}. RM09 list only five bona fide candidates for eclipsing Ap stars from which one, AO Velorum, has been confirmed \citep{Gonz06}. 
In their photometric study of Ap stars with the STEREO satellites, \citet{Wrai12} did not identify any eclipsing binaries or ellipsoidal variables 
among their sample of 337 stars. However, some of them show distinct double-waved light curves. Furthermore, for the spectroscopic binaries with 
known periods among their sample, they were not able to establish a relation between the orbital period and the period of the observed light variations.

To sum up, it seems to be well established that the incidence of ellipsoidal variables or eclipsing binaries among Ap stars is very low. Accordingly, 
if the observed variability pattern is in general accordance with the rotational modulation caused by spots, we are inclined to interpret 
double-waved light curves as due to this mechanism. However, in doubtful cases, orbital motion cannot be decisively rejected as the source of the observed 
variability.

In fact, we have identified some stars among our sample that we consider to be promising eclipsing binary or ellipsoidal variable candidates, 
which -- considering the size of the sample -- is expected. HD 70817 is very likely an eclipsing binary, but its status as a 
CP star seems to be doubtful \citep{Skif14}. Multicolour photometry with higher precision and/or 
spectroscopic studies are needed for a final conclusive classification. This will be part of a follow-up investigation.

\begin{table*}
\caption{Essential data for the 7 stars identified as promising eclipsing binary or ellipsoidal variable star candidates in the present paper.}
\label{table_2a}
\scriptsize{
\begin{center}
\begin{tabular}{llcclcccccc}
\hline
\hline 
Star &  ID (RM09)       & $\alpha$ (J2000) & $\delta$ (J2000)   & Type & Range ($V$) & Period & Epoch (HJD)              & Spectral type &       $(B-V)$ & $(J-K_{\mathrm s})$ \\
     &              &                  &                  &      & [mag]       & [d]    & [d]                 &               & [mag]   & [mag] \\
\hline
HD 57526        &       15710   &       07 20 52.724    &       $-$21 34 30.25   &       ACV/EB  &       8.33$-$8.37     &       3.3446(2)       &       2452702.65(7), min     &       B9 Si   &       $-$0.040        &       $-$0.035        \\
HD 64881        &       17800   &       07 53 59.051    &       $-$45 38 35.78   &       ACV/ELL &       8.79$-$8.83     &       1.1039(2)       &       2453057.62(2), max     &       A0 Si   &       $-$0.033        &       $-$0.024        \\
HD 70817        &       19540   &       08 22 33.328    &       $-$39 28 23.44   &       EB      &       9.30$-$9.45     &       3.50859(1)      &       2454181.730(4), min     &       B9 Si   &       +0.201  &        +0.194 \\
HD 97986        &       28260   &       11 14 49.418    &       $-$70 21 37.66   &       ACV/ELL &       7.84$-$7.90     &       4.3538(3)       &       2451878.79(8), max     &       B8 Si   &       +0.046  &       $-$0.011        \\
HD 133246       &       37830   &       15 05 30.511    &       $-$53 31 53.40   &       ACV/ELL &       9.05$-$9.10     &       1.79858(4)      &       2451953.79(4), max     &       B9 Si   &       +0.041  &       $-$0.024        \\
HD 161724       &       45640   &       17 50 29.078    &       $-$56 34 29.24   &       ACV/EB  &       9.60$-$9.67     &       5.6882(4)       &       2452067.6(1), min     &       B9 Si   &       $-$0.039        &       $-$0.028        \\
HD 174595       &       48860   &       18 52 04.499    &       $-$20 36 17.10   &       ACV/EB  &       9.08$-$9.17     &       7.1142(7)       &       2452464.6(1), min     &       A0 Si   &       +0.095  &       $-$0.031        \\
\hline
\end{tabular}
\end{center}
}
\end{table*}

\section{Results} \label{results}

Most of the investigated Ap stars have never been the subject of a light variability analysis before, or, as in the case of 
HD~263361, they have been investigated and found constant or probably constant, and are described here as variable stars for the 
first time. Some of our targets have been identified as variable stars with or without a given period in the literature, but 
their variability types have not been determined or they have been misclassified. To the best of our knowledge, these stars are 
presented here as ACV variables for the first time. Interestingly, 16 of these objects already have known or suspected periods as 
listed in the RM09 catalogue. While this implies they are members of the class of ACV variables, these stars are not listed as 
known or suspected ACV variables in related papers and variability catalogues. Thus, we deemed it worthwhile to draw 
attention to these objects by including them in our sample.

Also included in our sample are three Am stars from the RM09 list, which exhibit light variations and periods typical of 
ACV variables. Two of them are marked as not likely to be CP objects in RM09. 
Under the assumption that these stars might possibly have been spectroscopically misclassified, we have listed them as 
ACV candidates (type `ACV:' in Table \ref{table_2}) and suggest further spectroscopic observations. The objects are flagged as 
Am stars in Table \ref{table_3}.

\begin{table*}[ht]
\caption{Available information on single objects from the literature and miscellaneous remarks. An asterisk in column 7 
 denotes stars whose status as chemically peculiar objects is doubtful according to RM09.
Only part of the table is printed here for guidance regarding its form and content. The complete table is given in the Appendix.}
\label{table_3}
\scriptsize{
\begin{center}
\begin{tabular}{llccccl}
\hline
\hline
Star &  Var. desig. & Var. type &       Period (d)  &   Period (d)      &       Reference       &       Remarks/comments        \\
     &  Literature       & Literature     & Literature  & This work   &           & Literature \\      
\hline
HD 30898        &       NSV 16178       &       VAR     &       3.61890 (VSX), 0.782 (RM09)    &       0.781776(8)     &       \citet{Leeu97}  &       RM09: Am star \\
HD 35177        &       HIP 25163       &       VAR     &       0.52820 (VSX)   &       0.528194(5)     &       \citet{Koen02}  &               \\
HD 36997        &       NSV 16417       &       VAR:    &       n/a     &       6.0072(4)       &       \citet{Olse83}  &               \\
HD 38366        &       ASAS J054533+0919.0     &       RRAB    &       0.62528 (VSX)   &       0.625264(5)     &       \citet{Pojm02}  &       RS12: ACV (prob: 0.7498)  \\
HD 39635        &       NSV 16723       &       VAR     &       3.96574 (VSX)   &       3.9660(2)       &       \citet{Leeu97}  &       R12: BE+GCAS/BE+GCAS (prob: 0.27/0.60)       \\
HD 44456        &       ASAS J062155+0018.3     &       ED, ESD &       7.78607 (VSX)   &       3.8934(2)       &       \citet{Pojm02}  &       RS12: ACV (prob 0.5627)   \\
HD 47714        &       HIP 31851       &       n/a     &       4.6946 (R12)    &       4.6953(4)       &       R12     &       R12: ACV/ACV (prob: 0.26/0.79)       \\
HD 263361       &               &               &               &       88.9(1) &               &       Constant or probably constant according \\
          &   &   &   &       &   & to \citet{Wrai12}; period is outside their \\
                                        &   &   &   &       &   & considered frequency domain        \\
HD 50391        &               &               &               &       1.18285(3)      &               &       *       \\
HD 50461        &       NSV 17218       &       VAR     &       0.89410 (VSX)   &       0.89403(2)      &       \citet{Leeu97}  &       R12: SPB/SPB (prob: 0.63/0.62)       \\
HD 50855        &       HIP 33125       &       n/a     &       2.24027 (R12)   &       2.23913(6)      &               &       R12: BE+GCAS/BE+GCAS (prob: 0.56/0.56)       \\
HD 50895        &       HIP 33164       &       VAR     &       3.58989 (VSX)   &       3.5902(2)       &       \citet{Koen02}  &       R12: BE+GCAS/ACV (prob: 0.44/0.56)   \\
HD 51031        &       ASAS J065504-1629.7     &       DCEP-FO &       2.364 (VSX)   &       2.36431(9)      &       \citet{Pojm02}  &       RS12: ACV (prob: 0.8802)  \\
HD 51303        &       ASAS J065645-0055.0     &       VAR     &       4.508258 (VSX)   &       4.5076(3)       &       \citet{Pojm00}  &               \\
HD 52567        &       LR CMa  &       ACYG:   &       11.942 (VSX)    &       11.9467(9)      &       \citet{Leeu97}  &       \citet{Duba11}: likely ACV; \\
          &         &       &               &         &                 & RS12: ACV (prob: 0.9641)        \\
HD 53204        &       NSV 17317       &       n/a     &       4.2462 (R12)    &       2.28534(7)      &       \citet{Leeu97}  &       R12: BE+GCAS/ACV (prob: 0.42/0.62)   \\
HD 53695        &       HIP 34175       &       VAR     &       0.09964 (VSX)   &       0.83891(2)      &       \citet{Koen02}  &       *       \\
HD 53851        &       HIP 34283       &       VAR     &       2.02446 (VSX), 4.049? (RM09)   &       2.02464(5)      &       \citet{Koen02}  &       R12: SPB/SPB (prob: 0.50/0.80)       \\
\hline
\end{tabular}
\end{center}
}
\end{table*}

Some stars in our sample are listed with a twice longer or a twice shorter period in the literature. A twice longer (or shorter) 
rotation period cannot, in some cases, be definitely rejected for objects showing sinusoidal light variations and very small 
amplitudes and/or significant scatter in their light curves. However, the observed discrepancy is mostly due to classification 
differences. For example, some ASAS variables, which we have identified as ACV variables, have been classified as eclipsing binaries in 
the ASAS Catalogue of Variable Stars (ACVS) and are consequently listed there with a  period that is twice as long. For very few cases, the 
period value listed in the literature is very different from our solution and possibly represents an alias of the real period.

We have checked the period solution of all doubtful cases (Sect. \ref{data_int}) and are confident that we have determined the period 
that fits ASAS-3 data best. This is also supported by the generally very good agreement of our period solutions to those from the 
literature. Because of the strong daily aliasing inherent to ASAS-3 data (Sect. \ref{ASAS3}), alias periods cannot be totally excluded.

Some stars in our sample ($N$\,=\,31) have been spectroscopically classified as late B/early A-type stars, but have not been identified 
as (possible) chemically peculiar objects in the literature. Nevertheless, they have been included as ACV candidates (denoted as type `ACV:' 
in Table \ref{table_2}) in the present list of stars because of their typical photometric variability. Spectroscopic confirmation of these stars 
is needed to draw a final conclusion about their nature.

\begin{table*}
\caption{Reddening values derived for 99 stars of our sample from $UBV$ and $uvby\beta$ photometry as well as the extinction model by \citet{Amor05};
$A_{\mathrm V}$ (=\,3.1$E(B-V)$\,=\,5.636$E(J-K_{\mathrm s})$).}
\label{reddening}
\begin{center}
\begin{tabular}{lclclclc}
\hline
\hline
Star & $E(B-V)$ & Star & $E(B-V)$ & Star & $E(B-V)$ & Star & $E(B-V)$ \\ 
     & [mag] & & [mag] & & [mag] & & [mag] \\ 
\hline
CD--44 3656     &       0.022   &       HD 53021        &       0.022   &       HD 88488   &       0.081   &       HD 138773       &       0.170   \\
CPD--60 944A    &       0.000   &       HD 53204        &       0.017   &       HD 88757   &       0.142   &       HD 144815       &       0.166   \\
HD 10081        &       0.015   &       HD 53695        &       0.069   &       HD 89519   &       0.000   &       HD 146555       &       0.138   \\
HD 30898        &       0.033   &       HD 53851        &       0.000   &       HD 92190   &       0.062   &       HD 148848       &       0.000   \\
HD 35177        &       0.040   &       HD 54544        &       0.025   &       HD 92379   &       0.000   &       HD 149831       &       0.178   \\
HD 36997        &       0.057   &       HD 56273        &       0.034   &       HD 92849   &       0.029   &       HD 154645       &       0.098   \\
HD 37189        &       0.046   &       HD 56748        &       0.055   &       HD 93821   &       0.007   &       HD 155313       &       0.252   \\
HD 37713        &       0.051   &       HD 56773        &       0.071   &       HD 94873   &       0.000   &       HD 157063       &       0.225   \\
HD 38698        &       0.046   &       HD 57372        &       0.029   &       HD 96910   &       0.000   &       HD 157644       &       0.121   \\
HD 39635        &       0.014   &       HD 57946        &       0.084   &       HD 98486   &       0.225   &       HD 157678       &       0.024   \\
HD 44290        &       0.002   &       HD 59437        &       0.029   &       HD 99204   &       0.096   &       HD 158596       &       0.049   \\
HD 44456        &       0.043   &       HD 59752        &       0.099   &       HD 103844  &       0.289   &       HD 161349       &       0.070   \\
HD 47714        &       0.056   &       HD 61622        &       0.054   &       HD 105457  &       0.207   &       HD 166198       &       0.041   \\
HD 48729        &       0.021   &       HD 62535        &       0.035   &       HD 106688  &       0.144   &       HD 169005       &       0.169   \\
HD 50221        &       0.028   &       HD 62632        &       0.032   &       HD 111055  &       0.011   &       HD 174646       &       0.000   \\
HD 50391        &       0.000   &       HD 64901        &       0.013   &       HD 112528  &       0.118   &       HD 180029       &       0.303   \\
HD 50461        &       0.000   &       HD 83002        &       0.041   &       HD 115789  &       0.000   &       HD 181550       &       0.038   \\
HD 50540        &       0.000   &       HD 84001        &       0.022   &       HD 119716  &       0.027   &       HD 187128       &       0.040   \\
HD 50895        &       0.000   &       HD 84448        &       0.058   &       HD 123627  &       0.107   &       HD 258583       &       0.043   \\
HD 51088        &       0.000   &       HD 84656        &       0.075   &       HD 123927  &       0.211   &       HD 289186       &       0.010   \\
HD 51172        &       0.011   &       HD 85453        &       0.068   &       HD 125903  &       0.047   &       HD 299070       &       0.287   \\
HD 51307        &       0.051   &       HD 85469        &       0.029   &       HD 128840  &       0.205   &       HD 305451       &       0.225   \\
HD 51426        &       0.053   &       HD 85629        &       0.035   &       HD 132988  &       0.021   &       HD 315873       &       0.508   \\
HD 51790        &       0.040   &       HD 88343        &       0.094   &       HD 138151  &       0.166   &       HD 342867       &       0.027   \\
HD 52589        &       0.000   &       HD 88385        &       0.053   &       HD 138519  &       0.229   &       \\
\hline
\end{tabular}
\end{center}
\end{table*}
        
A part of our sample comprises Hipparcos unsolved variables \citep{Koen02}, which have been automatically 
classified using random forests \citep{Duba11} and a multistage methodology based on Bayesian networks 
\citep[][R12 hereafter]{Rimo12}. Likewise, the probabilistic classification catalogue created by 
\citet[][`Machine-learned ASAS Classification Catalog (MACC)'; RS12 hereafter]{Rich12} lists classifications and 
their probabilities for the (mis- or unclassified) ASAS variables included in our sample. Since the results of the 
employed automatic classification routines regarding our sample of ACV variables might potentially be of interest in 
estimating their performance, we have included the findings of the above-mentioned investigations in the presentation of our 
results. For stars having an entry in R12, we list variability types predicted by both methods, following the order of the 
VizieR online catalogue: predicted type random forests/predicted type Bayesian networks (probability predicted by random 
forests/probability predicted by Bayesian networks).
It is interesting to note that some of our confirmed ACV variables are listed as doubtful cases in RM09. That these 
objects exhibit light variations typical of ACV variables is further evidence of their peculiar nature.

Table \ref{table_statistics} presents statistical information on the composition of the present sample, and 
Tables \ref{table_2} to \ref{table_3} list the results of the present investigation and are presented in their entirety in the Appendix. 
Tables \ref{table_2} and \ref{table_2a} present essential data for the 323 stars we have identified as photometrically variable magnetic 
Ap stars or star candidates. The tables are organized as follows:

\begin{itemize}
\item Column 1: Star name, HD number, or other conventional identification
\item Column 2: Identification number from RM09
\item Column 3: Right ascension (Tycho-2)
\item Column 4: Declination (Tycho-2)
\item Column 5: Variability type, $\alpha^{2}$ Canum Venaticorum (ACV), eclipsing binary (EB), and ellipsoidal (ELL)  
\item Column 6: $V$ magnitude range
\item Column 7: Period (d)
\item Column 8: Epoch (HJD). In Table \ref{table_2a}, this column also specifies the type of epoch (Time of minimum is 
given for likely eclipsing binary or ellipsoidal variable star candidates, time of maximum for likely ACV variables.)
\item Column 9: Spectral classification, as listed in RM09 and -- for the objects not contained in this 
list -- \citet[][HD 39635]{Busc98}, \citet[][HD 52567, HD 88488, HD 89726, HD 166417, and HD 177548]{Cann93}, 
\citet[][HD 61134, HD 77013, HD 82691, HD 169005, and HD 299070]{Ochs80}, \citet{Skif14}, and
\citet[][TYC 8690-1870-1]{Sund74}. It is noteworthy that, as in the 
original paper, the `p' denoting peculiarity has been omitted from the spectral classifications taken from RM09
\item Column 10: $(B-V)$ index, taken from \citet{Khar01}
\item Column 11: $(J-K_{\mathrm s})$ index, as derived from the 2MASS catalogue \citep{Skru06}.
\end{itemize}

Table \ref{table_3} lists available information from the literature and miscellaneous remarks on individual objects. 
It is organized as follows:

\begin{itemize}
\item Column 1: Star name, HD number or other conventional identification
\item Column 2: Variable star designation from the literature
\item Column 3: Variable star type from the literature
\item Column 4: Period (d) from the literature
\item Column 5: Period (d) from this paper
\item Column 6: Reference in which -- to the best of our knowledge -- the object has been announced as a variable star for the first time
\item Column 7: Remarks of a miscellaneous nature: an asterisk denotes stars, whose status as chemically peculiar objects is doubtful according to RM09.
\end{itemize}

The light curves of all objects, folded with the period listed in Tables \ref{table_2} and \ref{table_2a}, are presented in the Appendix.

\subsection{Statistical analyses}

Because the source of our investigation, the RM09 catalogue, is rather inhomogeneous, the present sample is not very suitable for a
statistical analysis. However, to compare our results with those of other, similar investigations, we have 
included some informative plots. In the following, we do not include objects whose status as ACV variables remains doubtful (Table \ref{table_2a}).

Figure \ref{distributions} investigates the distribution of the logarithmic rotational periods (upper panel) and the $V$ magnitude ranges (lower panel) of the photometrically variable Ap stars and Ap star candidates of our 
sample. Our results are in excellent agreement with the literature, illustrating 
the well-known peak around $\log P$\,=\,0.4\,days (compare, for example, Figure 7 of RM09) and the magnitude range of the variations \citep{Math85}. 
It is worthwhile pointing out, though, that the observed sharp decrease in numbers below a $V$ magnitude range $<$\,0.03\,mag is at least 
partially due to observational bias, since the ASAS-3 measurement uncertainties approach this value for the fainter objects, thus preventing the detection of 
very low-amplitude variables among the fainter stars. We want to stress that we might have missed some very low amplitude variations due to the selection 
and analysis process (Sect. \ref{data_int}). The rotational periods of Ap
stars are more or less restricted to a rather narrow range. In light of the oblique rotator model, this means that a stable
magnetic field configuration, which is needed to generate surface spots, can only survive for a distinct range of equatorial rotational velocities.
For a detailed analysis of this important topic, one needs masses, luminosities, ages, and the inclination angles of rotating stars.
This information cannot be straightforwardly determined from photometric data alone.

Figure \ref{bv_jk} (upper panels) shows the colour indices $(J-K_{\mathrm s})$ and $(B-V)$ versus the logarithmic rotational periods. It 
is a matter of debate and even some controversy in the literature whether there should be a correlation or not. While RM09 found no 
correlation between the periods and the $(B-V)$ colours for their extended sample (see their Figure 8), \citet{Miku09} reached the 
conclusion that cooler magnetic CP stars (such as SrCrEu stars) rotate more slowly (see their Figure 2). Of course, relying solely on colour information 
simplifies the problem dramatically. This approach does not take the luminosity into account, as well as the mass and thus evolutionary
effects. We investigated such a possible simple colour correlation by employing a linear regression. 
We chose to restrict the colour index range to the most densely populated region of the diagram, which 
corresponds to $(B-V)$\,$\le$\,0.25\,mag. This also allowed us a direct comparison with Figure 2 of \citet{Miku09}.
In agreement with the findings of these investigators, we found a possible weak correlation between the rotational 
period and $(B-V)$ index ($t$ parameter of 3.87), in the sense that cooler magnetic CP stars rotate more slowly.
The slope was estimated as 0.57(19).

However, the reddening for our sample cannot be neglected. Most of our stars are located in the Galactic disk with 
a Galactic latitude $\left| b \right|$\,$<$\,10\degr\ and 7\,$\la$\,$V$\,$\la$\,10\,mag. For an A0 star with $M_{\mathrm V}$\,$\sim$\,0\,mag,
this corresponds to distances between 250 and 1000\,pc. In the Galactic disk, an absorption of $A_{\mathrm V}$ (=\,3.1$E(B-V)$\,=\,5.636$E(J-K_{\mathrm s})$) of
about 2\,mag\,kpc$^{-1}$ was observed for some regions \citep{Chen98,Dutr02}.

The reddening for the targets was estimated using photometric calibrations in the
Str{\"o}mgren $uvby\beta$ \citep{Craw78,Craw79} and the $Q$-parameter within
the Johnson $UBV$ system \citep{John58}. These methods are
only based on photometric indices and do not take
any distance estimates via parallax measurements into account. Photometric data were taken from the 
General Catalogue of Photometric Data
(GCPD\footnote{http://obswww.unige.ch/gcpd/}). Where possible, averaged and
weighted mean values were used throughout. For stars with Hipparcos parallax measurements \citep{Leeu07}
with a precision better than 30\%, the distance and Galactic coordinates were used to determine
the reddening from the extinction model by \citet{Amor05}. If several estimates were available, they were
compared and yielded excellent agreement. Table \ref{reddening} lists the reddening values of the 99
stars for which we are able to determine the reddening. For the remaining stars neither photometry nor
accurate parallaxes are available. The values for $A_{\mathrm V}$ reach up to 1.5\,mag.

As described above, we then employed a linear regression to the dereddened sample (Fig. \ref{bv_jk}, lower panels). No significant slope was 
found for this data sample. This could be interpreted as introducing a severe bias in the correlation when neglecting the reddening.
To exclude the possibility of a selection effect, we also investigated the reddened colours of the sample from Table \ref{reddening}.

Although the slopes are slightly shallower, they are still significant (Fig. \ref{bv_jk}, middle panels).
To verify our results, we investigated the behaviour of the correlation using Tycho-2 data instead of \citet{Khar01} data, 
which led to the same conclusions. Furthermore, we employed the Spearman rank-order correlation coefficient \citep{Spea04}, a non-parametric measure of correlation, to investigate our results. For this, we divided our sample (Table \ref{table_2}) into four groups using $\log P$ and $(B-V)$:
\begin{itemize}
\item All stars, reddened: $\rho$\,=\,+0.20, $p$\,=\,0.0004, $N$\,=\,316;
\item All stars with $(B-V)$\,$\le$\,0.25\,mag, reddened: $\rho$\,=\,+0.21, $p$\,=\,0.0002, $N$\,=\,295;
\item All stars, unreddened: $\rho$\,=\,$-$0.05, $p$\,=\,0.59, $N$\,=\,99;
\item All stars, with $(B-V)_{\mathrm 0}$\,$\le$\,0.25\,mag, unreddened: $\rho$\,=\,$-$0.03, $p$\,=\,0.76, $N$\,=\,98.
\end{itemize}
The two reddened samples show, with almost 100\% probability, that there is a weak correlation that is not
due to random sampling. In contrast, the other two samples are, with high probability, uncorrelated.
Therefore, from a statistical point of view, the derived coefficients substantiate our prior conclusions.
Clearly, a larger sample of CP stars with available rotational periods, along with precise stellar  
astrophysical parameters, is needed to investigate this topic further.

\section{Conclusions} \label{conclusions}

We have searched for photometric variability in confirmed or suspected Ap stars from the most recent version of the 
Catalogue of Ap, HgMn, and Am stars (RM09) using publicly available observations from the ASAS-3 database. We presented a 
sample of 323 photometrically variable magnetic Ap stars or star candidates, 246 of which are -- to the best of our 
knowledge -- described here as variable stars for the first time. A part of our sample (31 stars) is made up of 
late B to early A-type variables that are not included in the RM09 catalogue and that lack confirmation of their chemical peculiarity. 
They were included as ACV candidates because of their typical photometric variability but need spectroscopic confirmation.

We have compared our sample with the findings of similar investigations and find very good agreement. The well-known peak 
around $P$\,=\,2\,days in the distribution of rotational periods was confirmed. We found a possible weak correlation between 
rotational period and $(B-V)$ index, in that cooler magnetic CP stars rotate more slowly. However, this 
correlation seems to disappear when correcting for the interstellar reddening.

In agreement with the findings of \citet{Davi14} and \citet{Pigu14}, we conclude that ASAS-3 data are well suited 
to investigating variable stars with rather low photometric amplitudes.

\begin{acknowledgements}
This project is financed by the SoMoPro II programme (3SGA5916). The research leading
to these results received a financial grant from the People Programme
(Marie Curie action) of the Seventh Framework Programme of the EU according to REA Grant
Agreement No. 291782. The research is further co-financed by the South-Moravian Region. 
It was also supported by grant 7AMB14AT015 and
the financial contributions of the Austrian Agency for International 
Cooperation in Education and Research (BG-03/2013 and CZ-09/2014). 
We thank the referee, Luca Fossati, for helpful comments and suggestions that helped to improve the paper.
This work reflects only the author's views so the European 
Union is not liable for any use that may be made of the information contained therein.
\end{acknowledgements}

\appendix

\section{Complete Table 2}      

\begin{table*}
\caption{Essential data for the 316 stars identified as photometrically variable chemically peculiar stars or star candidates in the present paper.}
\scriptsize{
\begin{center}
\begin{tabular}{llcclcccccc}
\hline
\hline 
Star &  ID (RM09)       & $\alpha$ (J2000) & $\delta$ (J2000)   & Type & Range ($V$) & Period & Epoch (HJD)      & Spectral type &       $(B-V)$ & $(J-K_{\mathrm s})$ \\
     &              &                  &                  &      & [mag]       & [d]    & [d]         &               & [mag]   & [mag] \\
\hline
HD 10081        &       2510    &       01 36 15.100    &       $-$68 15 05.38   &       ACV     &       9.60$-$9.64     &       1.57032(3)      &       2452190.56(3)   &       A0 Sr Eu   &       +0.031  &        +0.041 \\
HD 26726        &       6830    &       04 12 46.635    &       $-$24 47 48.55   &       ACV     &       9.81$-$9.90     &       5.3816(4)       &       2451902.6(1)    &       A2 Sr      &       +0.325  &        +0.006 \\
HD 30898        &       7983    &       04 51 28.382    &       $-$00 32 47.85   &       ACV:    &       7.84$-$7.88     &       0.781776(8)     &       2452185.79(2)   &       A1$-$F0 &       +0.269  &        +0.124  \\
HD 35177        &       8980    &       05 23 01.934    &       +01 41 48.94    &       ACV     &       8.12$-$8.15     &       0.528194(5)     &       2452670.60(1)   &       B9 Si      &       $-$0.068        &       $-$0.118        \\
HD 36997        &       9810    &       05 35 13.819    &       $-$02 22 52.32   &       ACV     &       8.31$-$8.36     &       6.0072(4)       &       2453101.5(1)    &       B9 Si Sr   &       $-$0.043        &       $-$0.060        \\
HD 37189        &       9940    &       05 36 09.774    &       $-$13 49 20.35   &       ACV     &       8.53$-$8.57     &       2.8286(1)       &       2452115.90(6)   &       A0 Si      &       $-$0.033        &       $-$0.001        \\
\hline                                                                                                                                                                  
\end{tabular}                                                                                                                                                                   
\end{center}                                                                                                                                                                    
}                                                                                                                                                                       
\end{table*}                                                                                                                                                                    
\setcounter{table}{0}                                                                                                                                                                   
\begin{table*}                                                                                                                                                                  
\caption{continued.}                                                                                                                                                                    
\scriptsize{                                                                                                                                                                    
\begin{center}                                                                                                                                                                  
\begin{tabular}{llcclcccccc}                                                                                                                                                                    
\hline                                                                                                                                                                  
\hline                                                                                                                                                                  
Star &  ID (RM09)       & $\alpha$ (J2000) & $\delta$ (J2000)   & Type & Range ($V$) & Period & Epoch (HJD)      & Spectral type &       $(B-V)$ & $(J-K_{\mathrm s})$ \\                                                                                                                  
     &              &                  &                  &      & [mag]       & [d]    & [d]         &               & [mag]   & [mag] \\                                                                                                                                                                       
\hline                                                                                                                                                                  
HD 61334        &       16790   &       07 36 38.851    &       $-$45 16 29.10   &       ACV     &       9.17$-$9.20     &       19.150(5)       &       2452559.6(4)    &       A2 Si Cr   &       +0.026  &        +0.003 \\
HD 61134        &               &       07 36 56.131    &       $-$21 03 09.52   &       ACV:    &       8.48$-$8.52     &       4.6170(3)       &       2451968.65(9)   &       B9      &       $-$0.119        &       $-$0.104        \\
HD 61369        &       16820   &       07 38 08.275    &       $-$21 53 55.32   &       ACV     &       9.67$-$9.69     &       1.94372(5)      &       2454216.55(4)   &       B9 Si      &       $-$0.035        &       $-$0.063        \\
\hline                                                                                                                                                                  
\end{tabular}                                                                                                                                                                   
\end{center}                                                                                                                                                                    
}                                                                                                                                                                       
\end{table*}                                                                                                                                                                    
\setcounter{table}{0}                                                                                                                                                                   
\begin{table*}                                                                                                                                                                  
\caption{continued.}                                                                                                                                                                    
\scriptsize{                                                                                                                                                                    
\begin{center}                                                                                                                                                                  
\begin{tabular}{llcclcccccc}                                                                                                                                                                    
\hline                                                                                                                                                                  
\hline                                                                                                                                                                  
Star &  ID (RM09)       & $\alpha$ (J2000) & $\delta$ (J2000)   & Type & Range ($V$) & Period & Epoch (HJD)      & Spectral type &       $(B-V)$ & $(J-K_{\mathrm s})$ \\                                                                                                                  
     &              &                  &                  &      & [mag]       & [d]    & [d]         &               & [mag]   & [mag] \\                                                                                                                                                                       
\hline                                                                                                                                                                  
HD 82917        &       23620   &       09 33 25.835    &       $-$48 04 13.34   &       ACV     &       9.73$-$9.76     &       2.2460(1)       &       2453861.60(4)   &       A2 Si      &       $-$0.051        &       $-$0.045        \\
HD 83002        &       23670   &       09 33 36.434    &       $-$53 20 39.35   &       ACV     &       9.05$-$9.09     &       1.55104(3)      &       2454259.56(3)   &       B8 Si      &       $-$0.073        &       $-$0.091        \\
HD 82944        &       23630   &       09 34 44.056    &       $-$12 36 25.50   &       ACV     &       8.64$-$8.67     &       2.47906(8)      &       2454959.59(5)   &       A0 Sr Eu   &       +0.258  &       $-$0.008        \\
\hline                                                                                                                                                                  
\end{tabular}                                                                                                                                                                   
\end{center}                                                                                                                                                                    
}                                                                                                                                                                       
\end{table*}                                                                                                                                                                    
\setcounter{table}{0}                                                                                                                                                                   
\begin{table*}                                                                                                                                                                  
\caption{continued.}                                                                                                                                                                    
\scriptsize{                                                                                                                                                                    
\begin{center}                                                                                                                                                                  
\begin{tabular}{llcclcccccc}                                                                                                                                                                    
\hline                                                                                                                                                                  
\hline                                                                                                                                                                  
Star &  ID (RM09)       & $\alpha$ (J2000) & $\delta$ (J2000)   & Type & Range ($V$) & Period & Epoch (HJD)      & Spectral type &       $(B-V)$ & $(J-K_{\mathrm s})$ \\                                                                                                                  
     &              &                  &                  &      & [mag]       & [d]    & [d]         &               & [mag]   & [mag] \\                                                                                                                                                                       
\hline                                                                                                                                                                  
HD 132988       &       37760   &       15 05 24.700    &       $-$65 05 27.92   &       ACV     &       10.17$-$10.24   &       2.56368(8)      &       2452548.50(5)   &       A0 Si      &       +0.114  &        +0.047 \\
HD 133722       &       37930   &       15 07 49.229    &       $-$51 08 39.37   &       ACV     &       9.67$-$9.73     &       2.08809(6)      &       2452081.56(4)   &       B8 Si      &       +0.024  &       + 0.015 \\
HD 133702       &       37910   &       15 11 07.558    &       $-$73 08 18.19   &       ACV     &       9.48$-$9.54     &       2.10997(6)      &       2452102.49(4)   &       B8 Si      &       $-$0.069        &       $-$0.066        \\
\hline
\end{tabular}
\end{center}
}
\end{table*}

\section{Complete Table 4}      

\begin{sidewaystable*}
\centering
\caption{Available information on single objects from the literature and miscellaneous remarks. An asterisk in column 7 
(`Remarks/comments from literature') denotes stars, which status as chemically peculiar objects is doubtful according to RM09.}
\scriptsize{
\begin{tabular}{llccccl}
\hline
\hline
Star &  Var. desig. & Var. type &       Period (d)  &   Period (d)      &       Reference       &       Remarks/comments        \\
     &  Literature       & Literature     & Literature  & This work   &           & Literature \\      
\hline
HD 30898        &       NSV 16178       &       VAR     &       3.61890 (VSX), 0.782 (RM09)    &       0.781776(8)     &       \citet{Leeu97}  &       RM09: Am star \\
HD 35177        &       HIP 25163       &       VAR     &       0.52820 (VSX)   &       0.528194(5)     &       \citet{Koen02}  &               \\
HD 36997        &       NSV 16417       &       VAR:    &       n/a     &       6.0072(4)       &       \citet{Olse83}  &               \\
HD 38366        &       ASAS J054533+0919.0     &       RRAB    &       0.62528 (VSX)   &       0.625264(5)     &       \citet{Pojm02}  &       RS12: ACV (prob: 0.7498)  \\
HD 39635        &       NSV 16723       &       VAR     &       3.96574 (VSX)   &       3.9660(2)       &       \citet{Leeu97}  &       R12: BE+GCAS/BE+GCAS (prob: 0.27/0.60)       \\
HD 44456        &       ASAS J062155+0018.3     &       ED, ESD &       7.78607 (VSX)   &       3.8934(2)       &       \citet{Pojm02}  &       RS12: ACV (prob 0.5627)   \\
\hline
\end{tabular}
}
\end{sidewaystable*}
\setcounter{table}{0}
\begin{sidewaystable*}
\centering
\caption{continued.}
\scriptsize{
\begin{tabular}{llccccl}
\hline
\hline
Star &  Var. desig. & Var. type &       Period (d)  &   Period (d)      &       Reference       &       Remarks/comments        \\
     &  Literature       & Literature     & Literature  & This work   &           & Literature \\      
\hline
HD 81877        &       ASAS J092711-4022.1     &       DCEP-FO &       3.2152 (VSX)   &       3.2151(2)       &       \citet{Pojm02}  &       RS12: ACV (prob: 0.8197)  \\
HD 84001        &       NSV 18243       &       n/a     &       0.49671 (R12)   &       0.540937(5)     &       \citet{Leeu97}  &       R12: BE+GCAS/SPB (prob: 0.30/0.47)   \\
HD 84448        &       NSV 18260       &       VAR     &       1.013014 (VSX)   &       1.01297(1)      &       \citet{Leeu97}  &       \citet{Duba11}: likely SPB      \\
HD 84690        &               &               &               &       8.1212  &               &       *       \\
HD 85469        &       HIP 48222       &       n/a     &       1.3116 (R12)    &       2.4305(1)       &       R12     &       R12: BE+GCAS/SPB (prob: 0.32/0.65)   \\
HD 85629        &       HIP 48253       &       n/a     &       0.86000 (R12)   &       1.72016(4)      &       R12     &       R12: SPB/SPB (prob: 0.33/0.79)       \\
\hline
\end{tabular}
}
\end{sidewaystable*}

\section{Light curves}

\begin{figure*}
\begin{center}
\includegraphics[width=1.0\textwidth,natwidth=1440,natheight=1728]{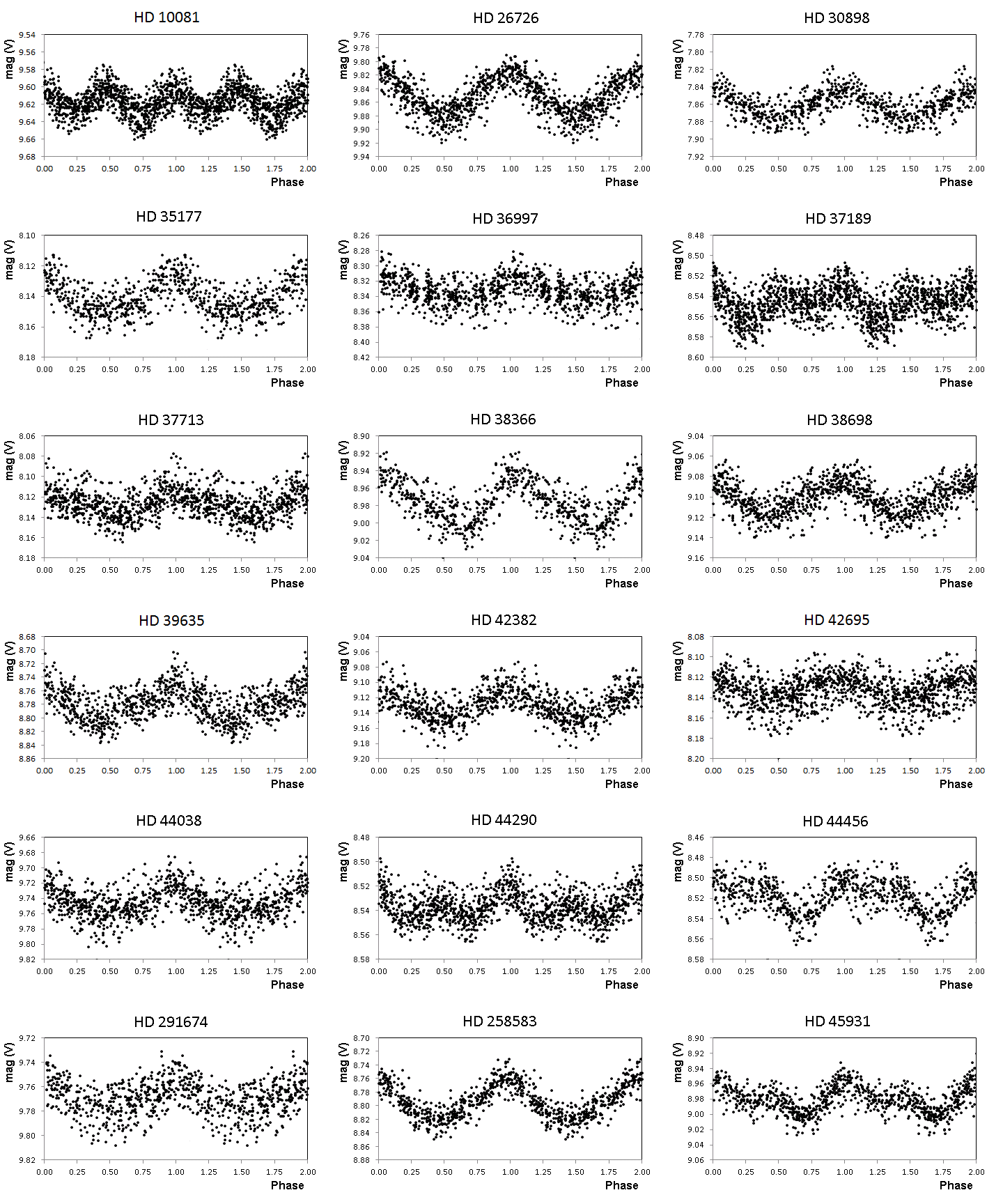}
\caption{The light curves of all objects, folded with the period period listed in Tables \ref{table_2} and \ref{table_2a}, respectively.} 
\end{center}
\end{figure*}
\setcounter{figure}{0}
\begin{figure*}
\begin{center}
\includegraphics[width=1.0\textwidth,natwidth=1440,natheight=1728]{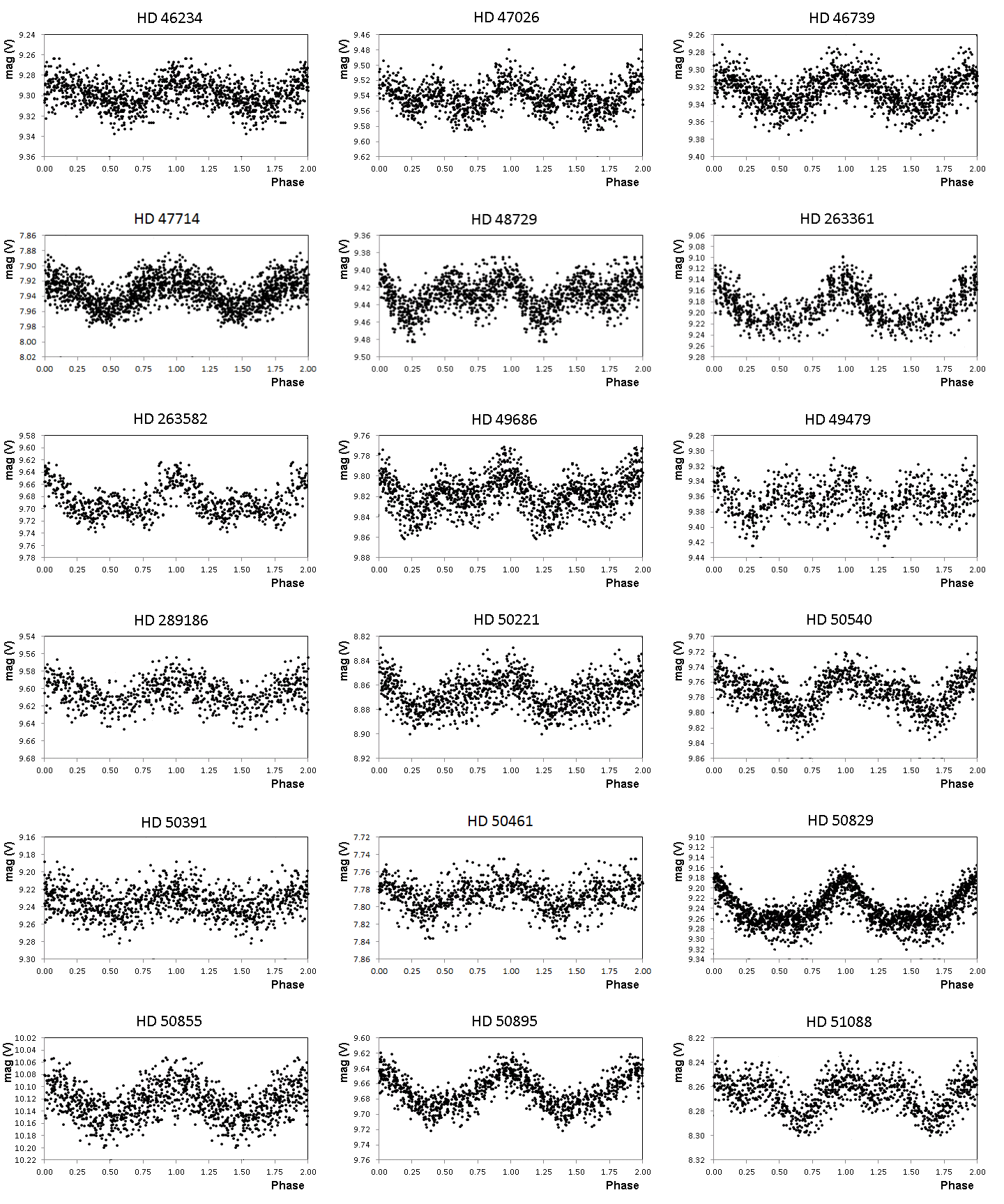}
\caption{continued.} 
\end{center}
\end{figure*}
\setcounter{figure}{0}
\begin{figure*}
\begin{center}
\includegraphics[width=1.0\textwidth,natwidth=1440,natheight=1728]{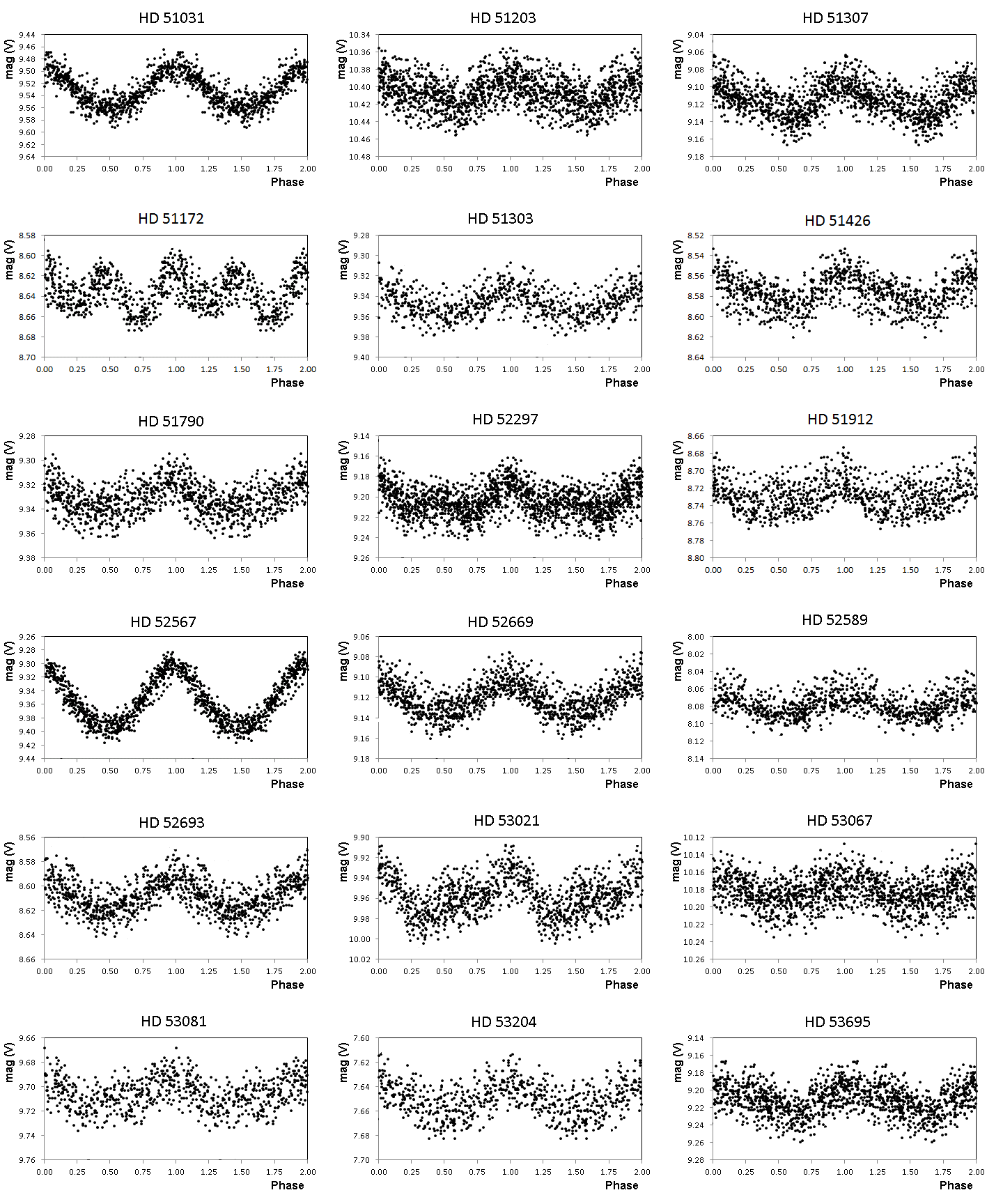}
\caption{continued.} 
\end{center}
\end{figure*}
\setcounter{figure}{0}
\begin{figure*}
\begin{center}
\includegraphics[width=1.0\textwidth,natwidth=1440,natheight=1728]{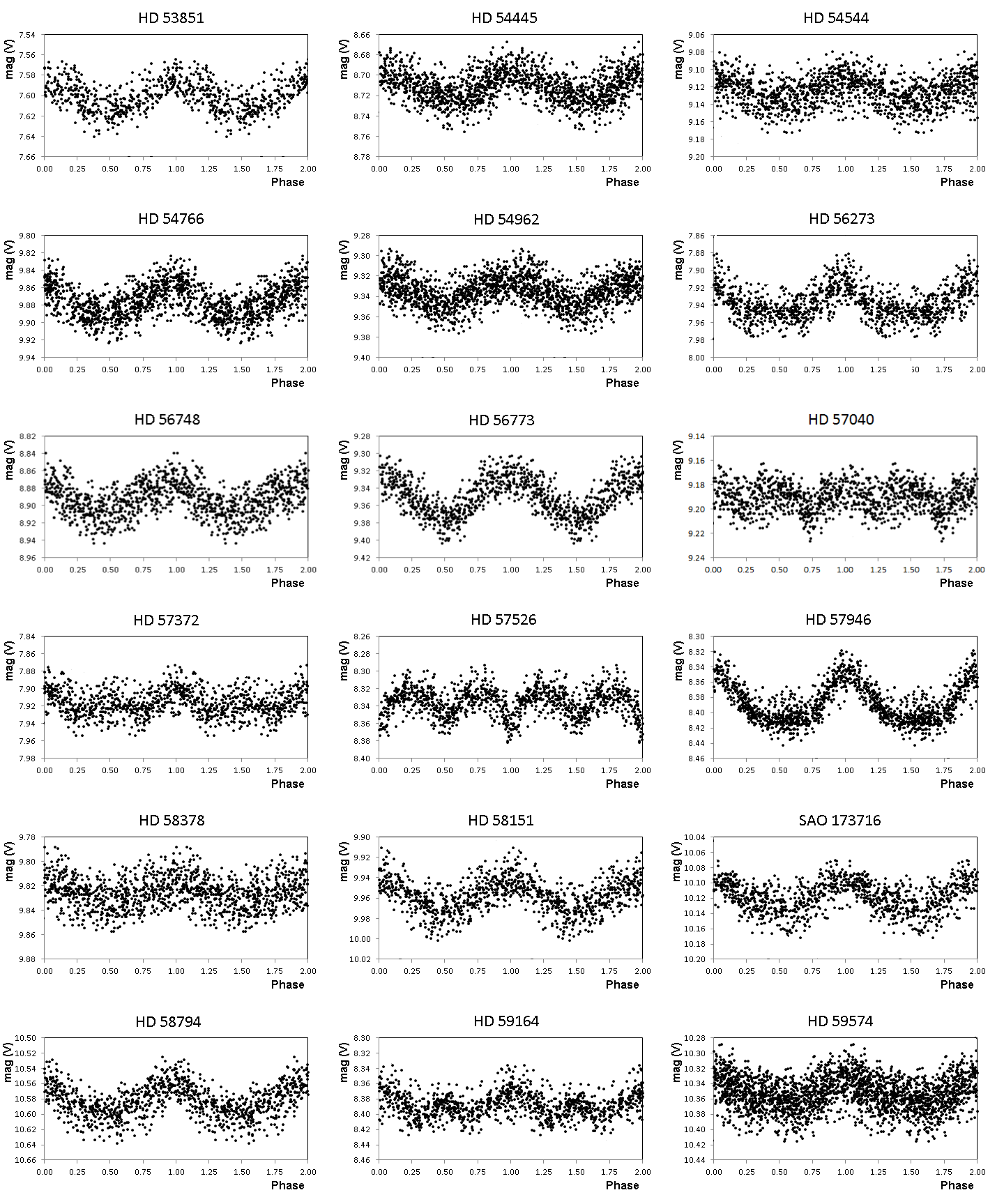}
\caption{continued.} 
\end{center}
\end{figure*}
\setcounter{figure}{0}
\begin{figure*}
\begin{center}
\includegraphics[width=1.0\textwidth,natwidth=1440,natheight=1728]{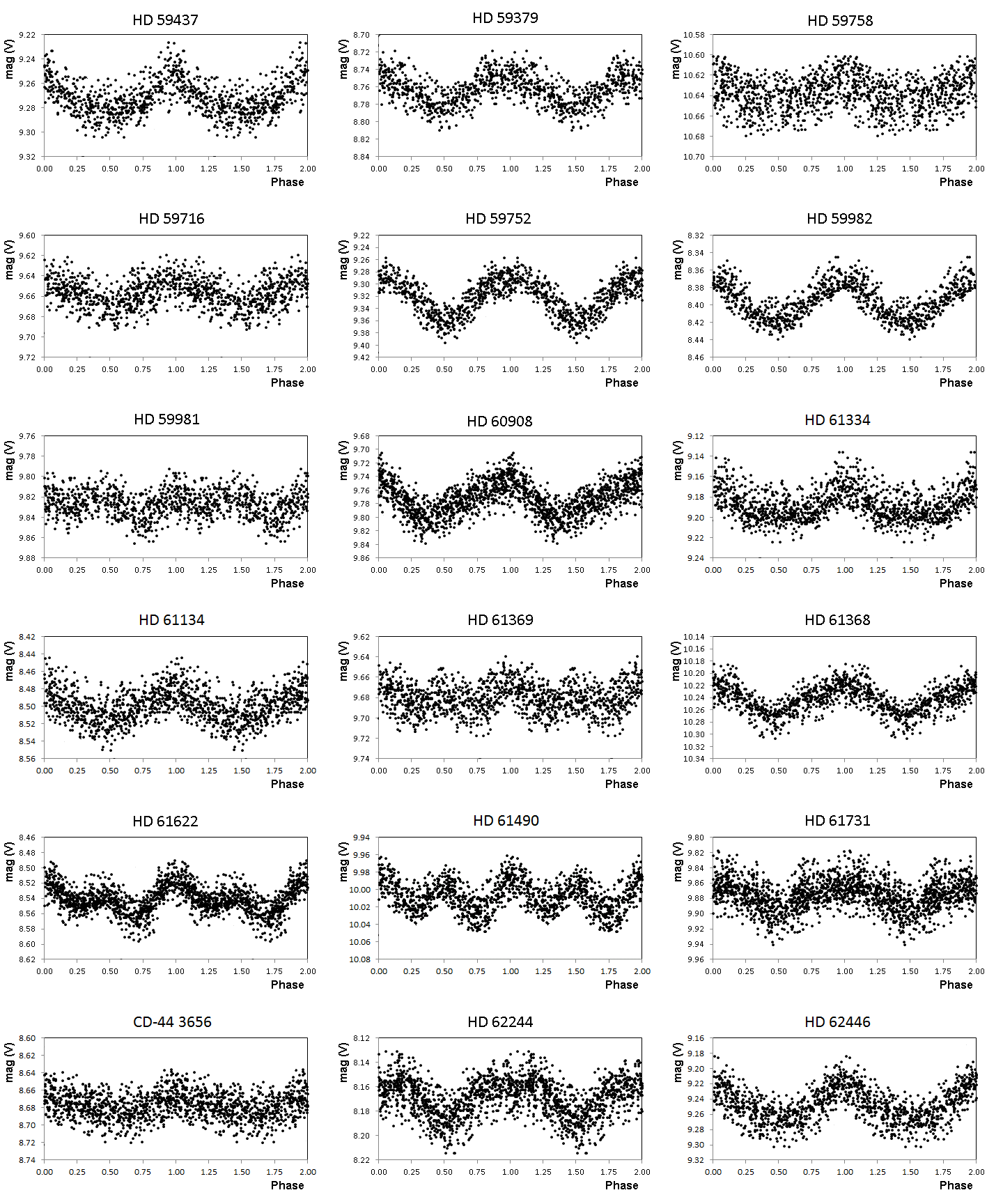}
\caption{continued.} 
\end{center}
\end{figure*}
\setcounter{figure}{0}
\begin{figure*}
\begin{center}
\includegraphics[width=1.0\textwidth,natwidth=1440,natheight=1728]{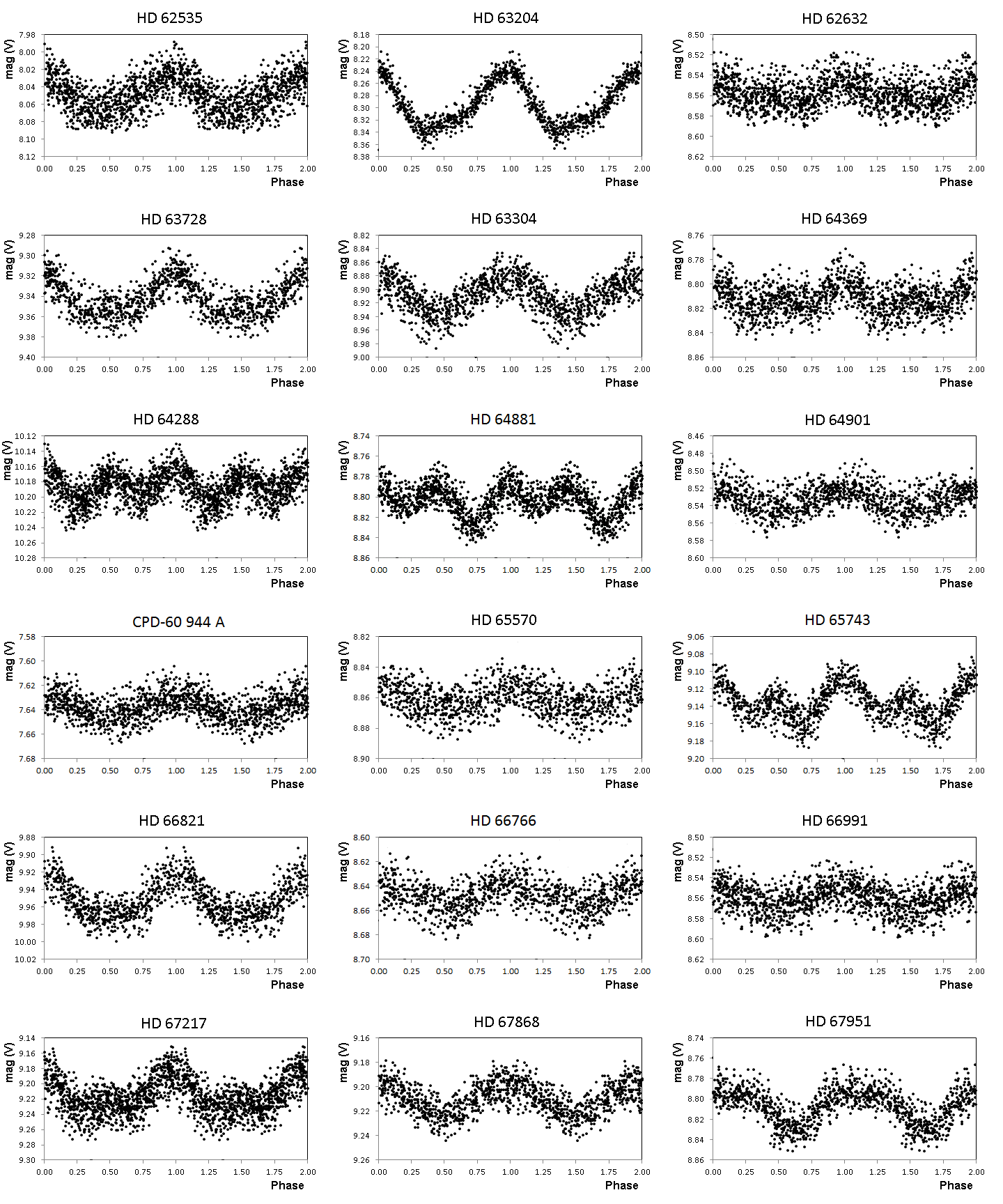}
\caption{continued.} 
\end{center}
\end{figure*}
\setcounter{figure}{0}
\begin{figure*}
\begin{center}
\includegraphics[width=1.0\textwidth,natwidth=1440,natheight=1728]{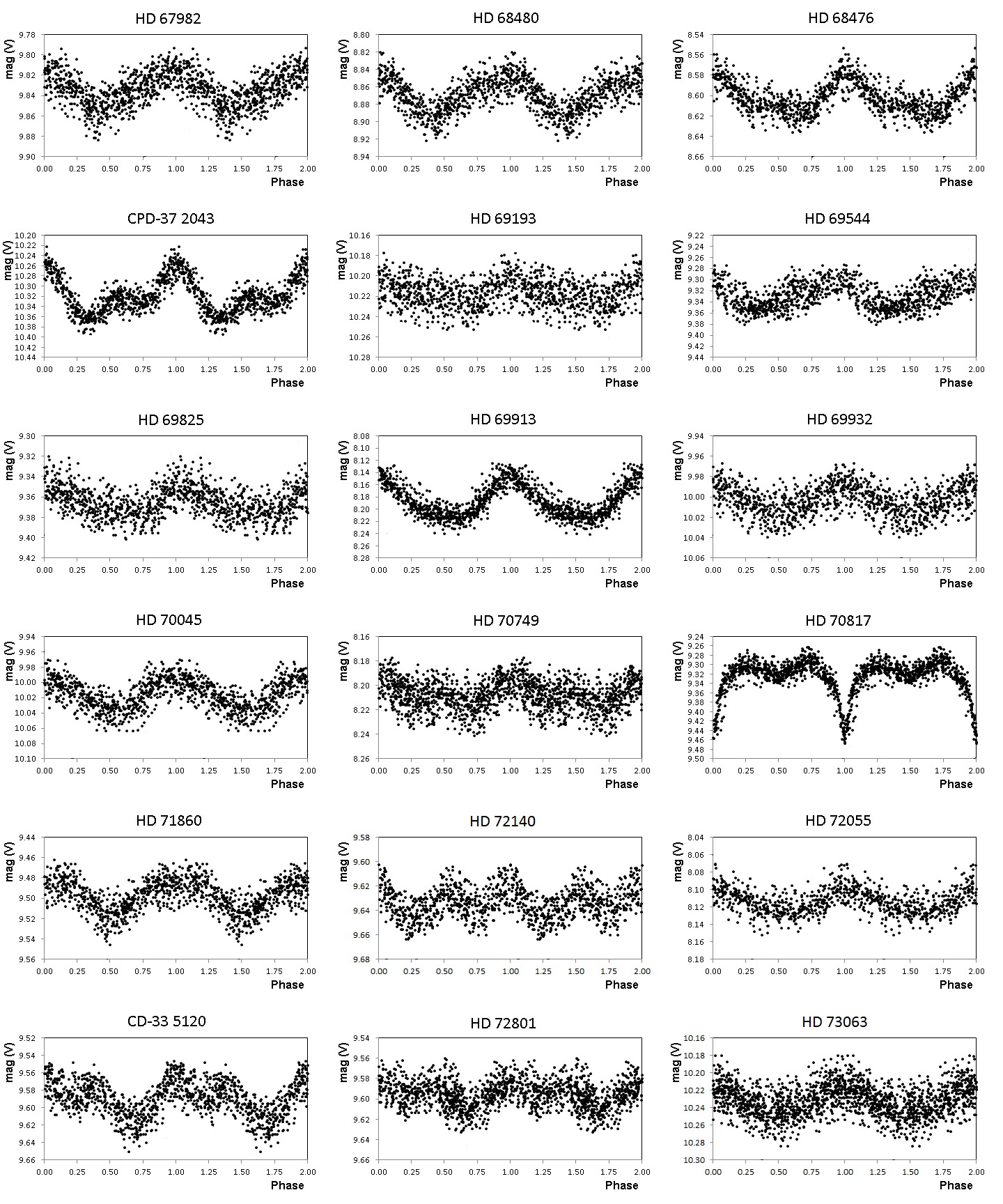}
\caption{continued.} 
\end{center}
\end{figure*}
\clearpage 
\setcounter{figure}{0}
\begin{figure*}
\begin{center}
\includegraphics[width=1.0\textwidth,natwidth=1440,natheight=1728]{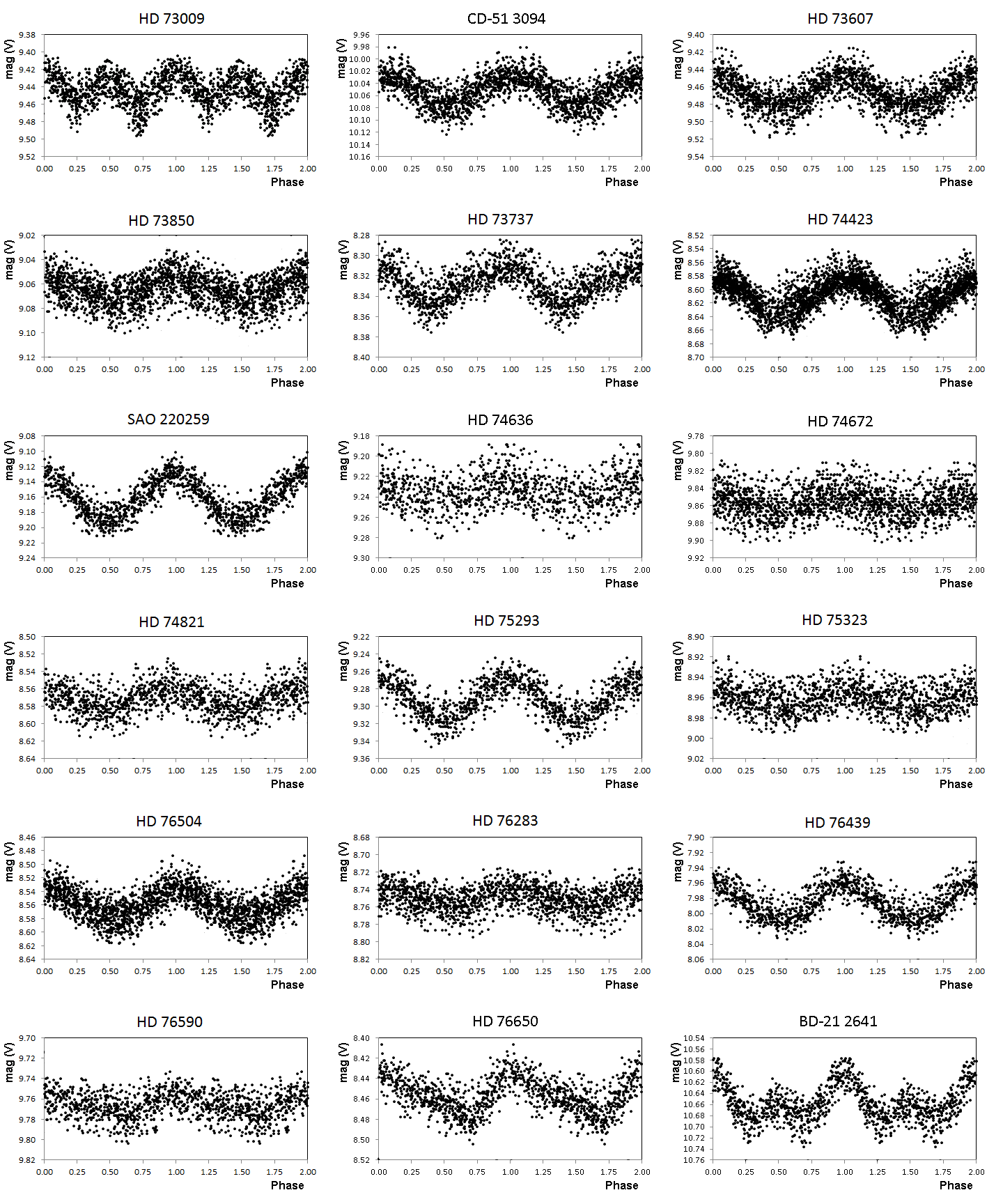}
\caption{continued.} 
\end{center}
\end{figure*}
\setcounter{figure}{0}
\begin{figure*}
\begin{center}
\includegraphics[width=1.0\textwidth,natwidth=1440,natheight=1728]{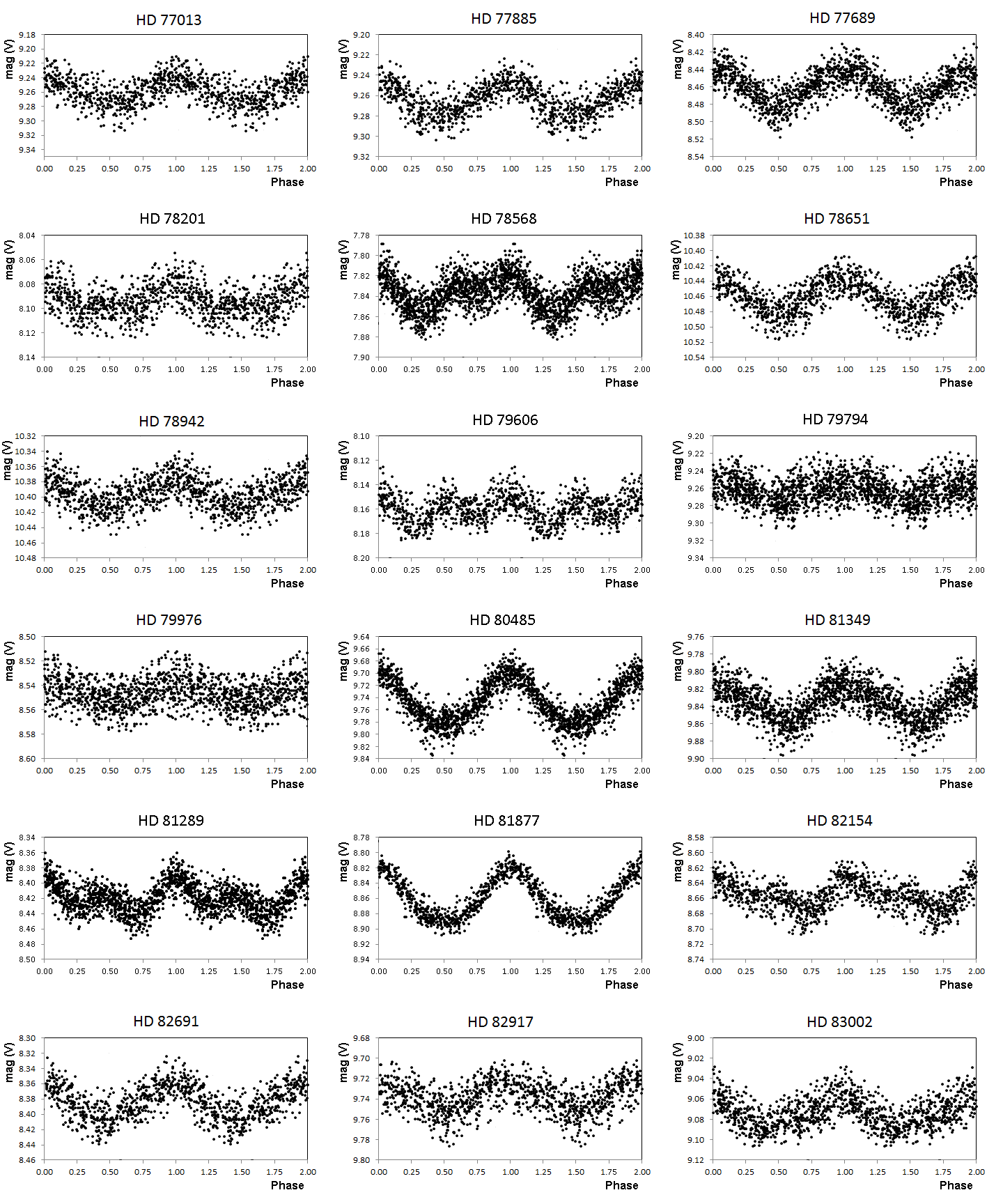}
\caption{continued.} 
\end{center}
\end{figure*}
\setcounter{figure}{0}
\begin{figure*}
\begin{center}
\includegraphics[width=1.0\textwidth,natwidth=1440,natheight=1728]{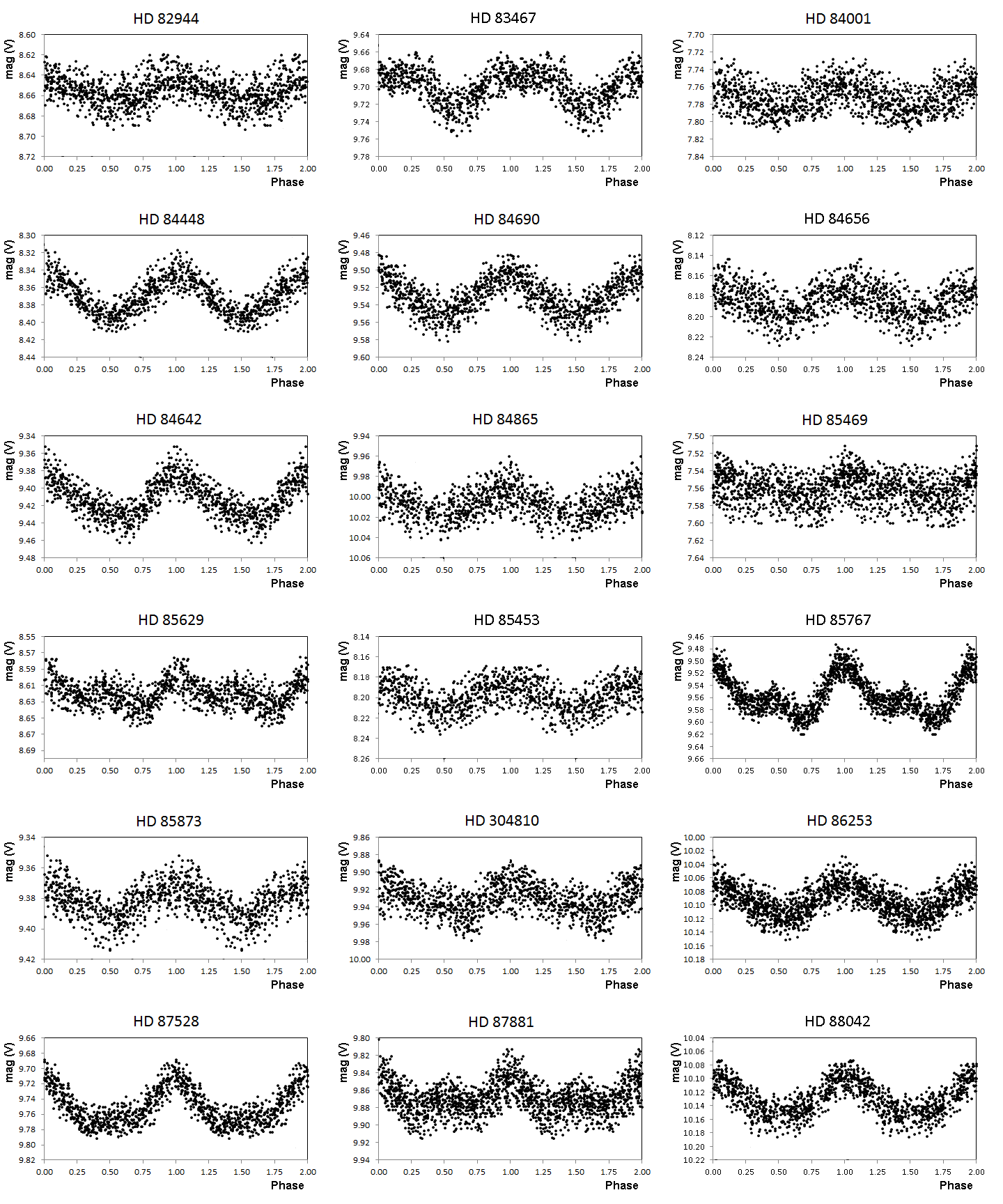}
\caption{continued.} 
\end{center}
\end{figure*}
\setcounter{figure}{0}
\begin{figure*}
\begin{center}
\includegraphics[width=1.0\textwidth,natwidth=1440,natheight=1728]{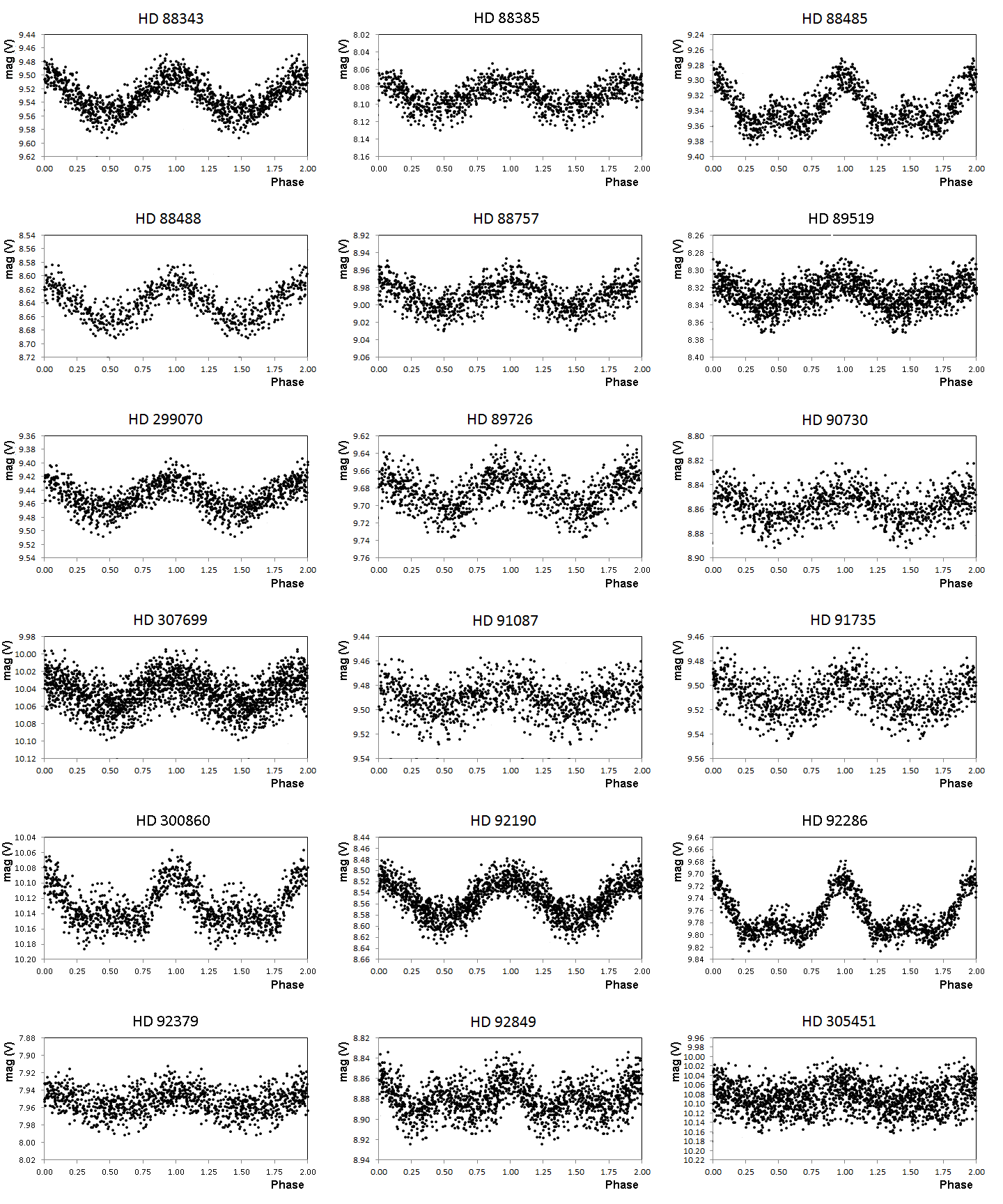}
\caption{continued.} 
\end{center}
\end{figure*}
\setcounter{figure}{0}
\begin{figure*}
\begin{center}
\includegraphics[width=1.0\textwidth,natwidth=1440,natheight=1728]{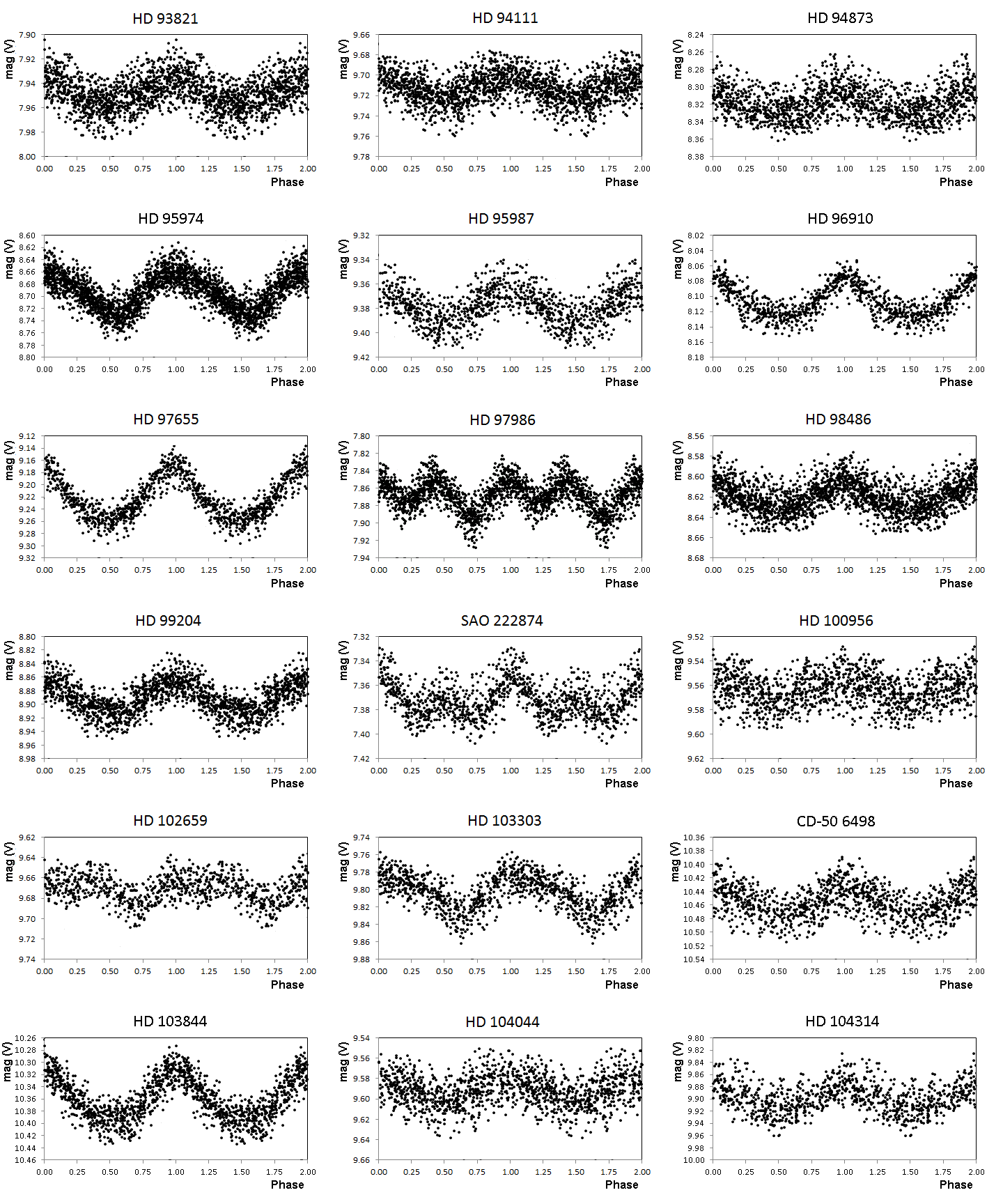}
\caption{continued.} 
\end{center}
\end{figure*}
\setcounter{figure}{0}
\begin{figure*}
\begin{center}
\includegraphics[width=1.0\textwidth,natwidth=1440,natheight=1728]{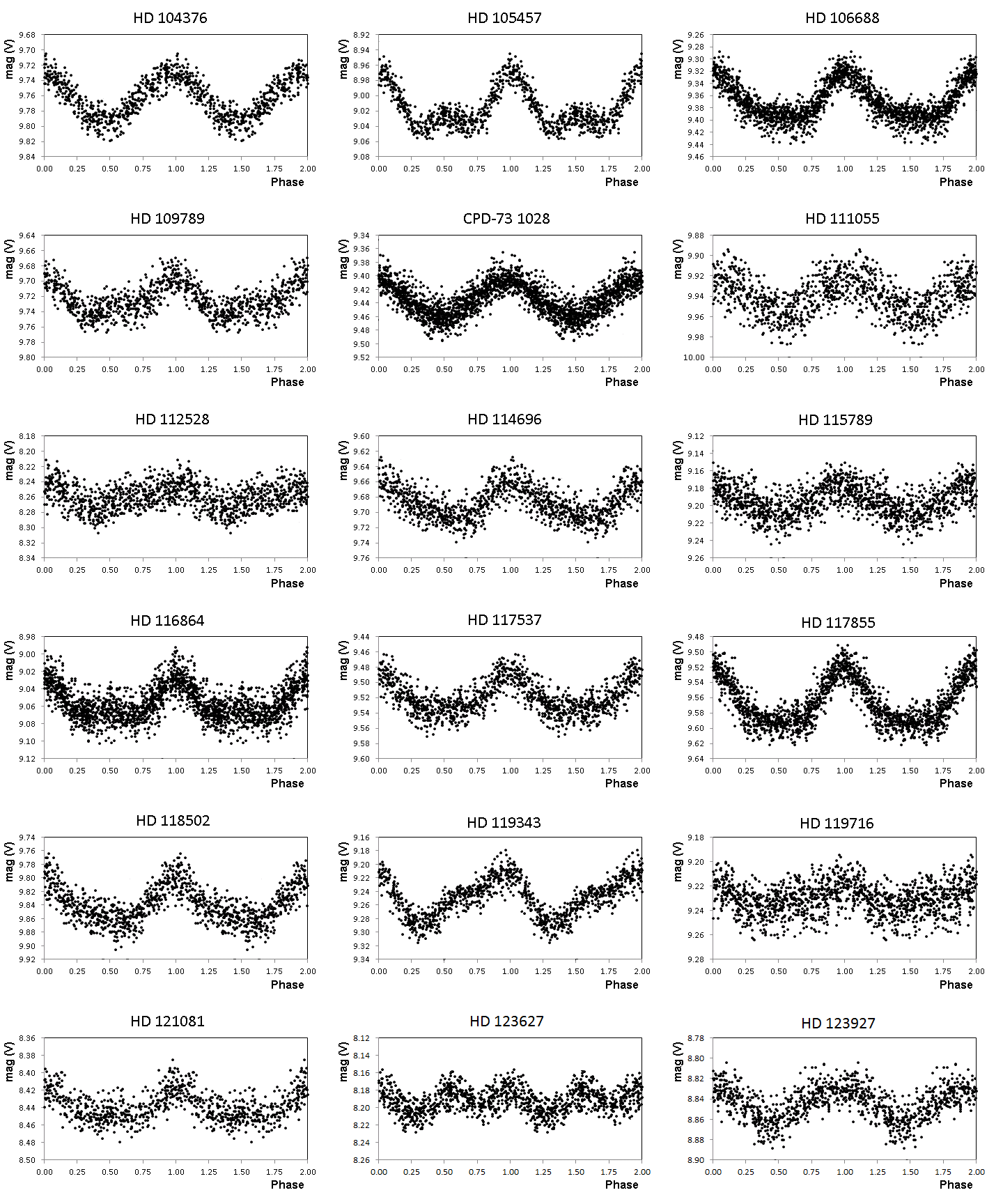}
\caption{continued.} 
\end{center}
\end{figure*}
\setcounter{figure}{0}
\begin{figure*}
\begin{center}
\includegraphics[width=1.0\textwidth,natwidth=1440,natheight=1728]{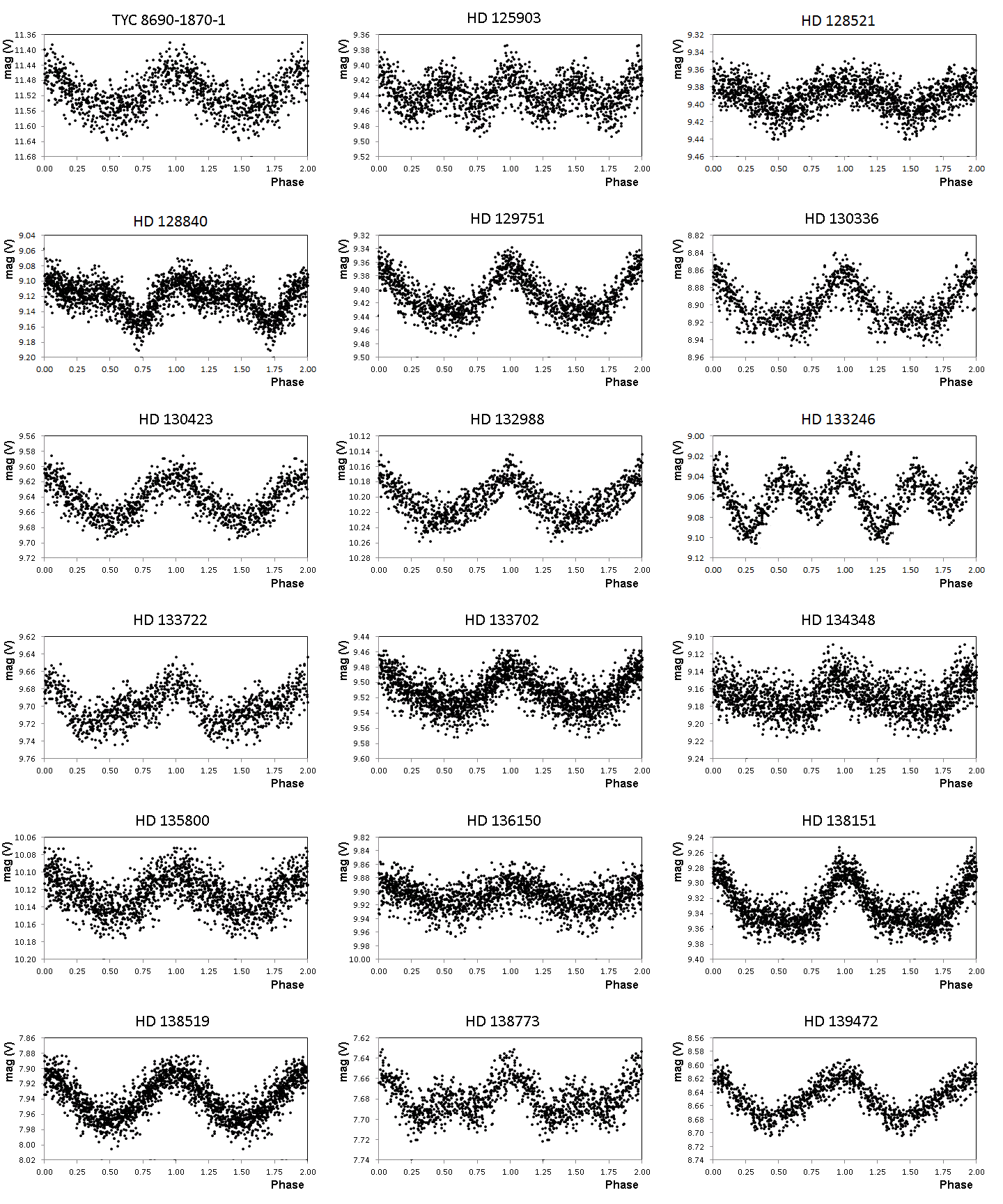}
\caption{continued.} 
\end{center}
\end{figure*}
\setcounter{figure}{0}
\begin{figure*}
\begin{center}
\includegraphics[width=1.0\textwidth,natwidth=1440,natheight=1728]{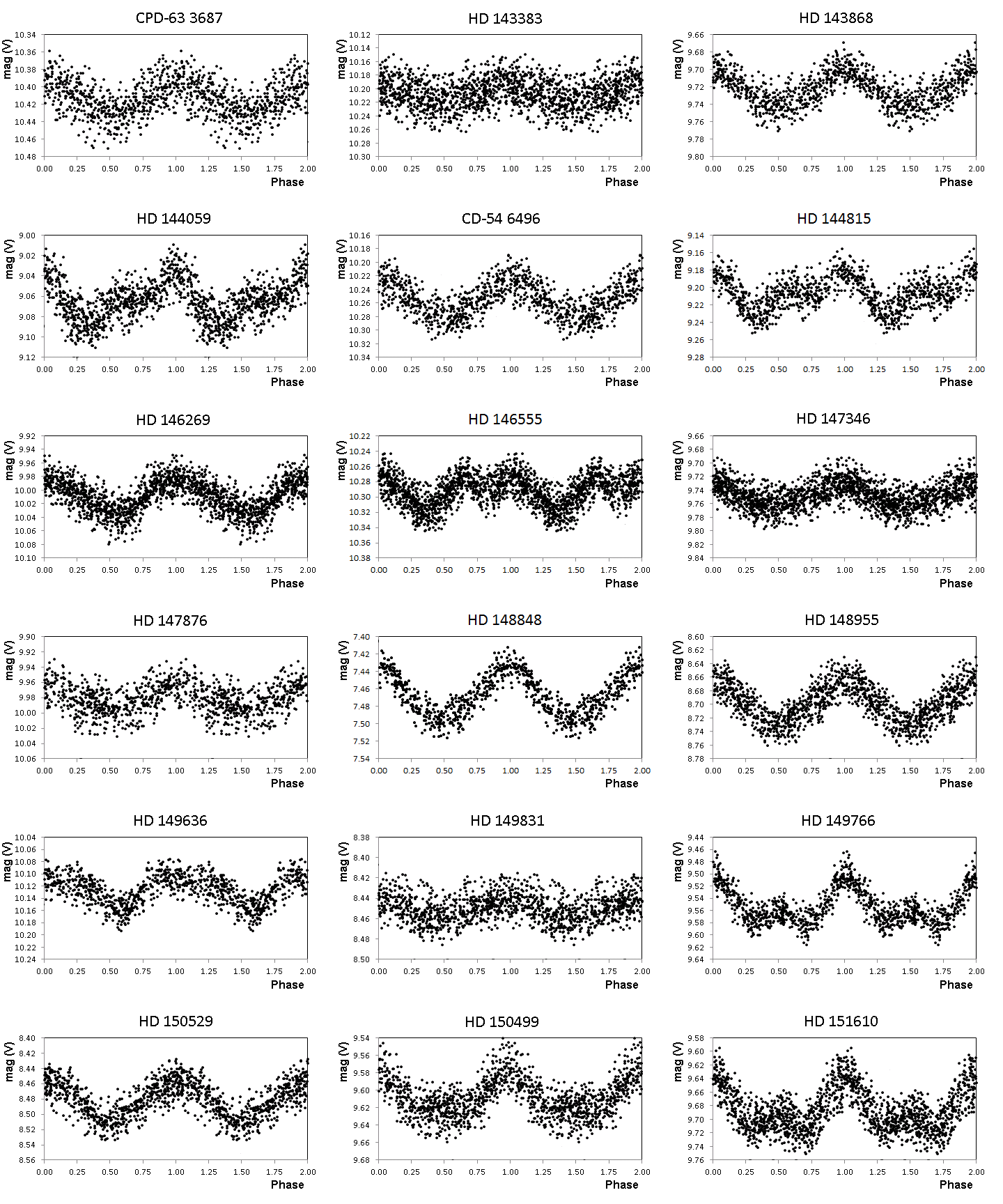}
\caption{continued.} 
\end{center}
\end{figure*}
\setcounter{figure}{0}
\begin{figure*}
\begin{center}
\includegraphics[width=1.0\textwidth,natwidth=1440,natheight=1728]{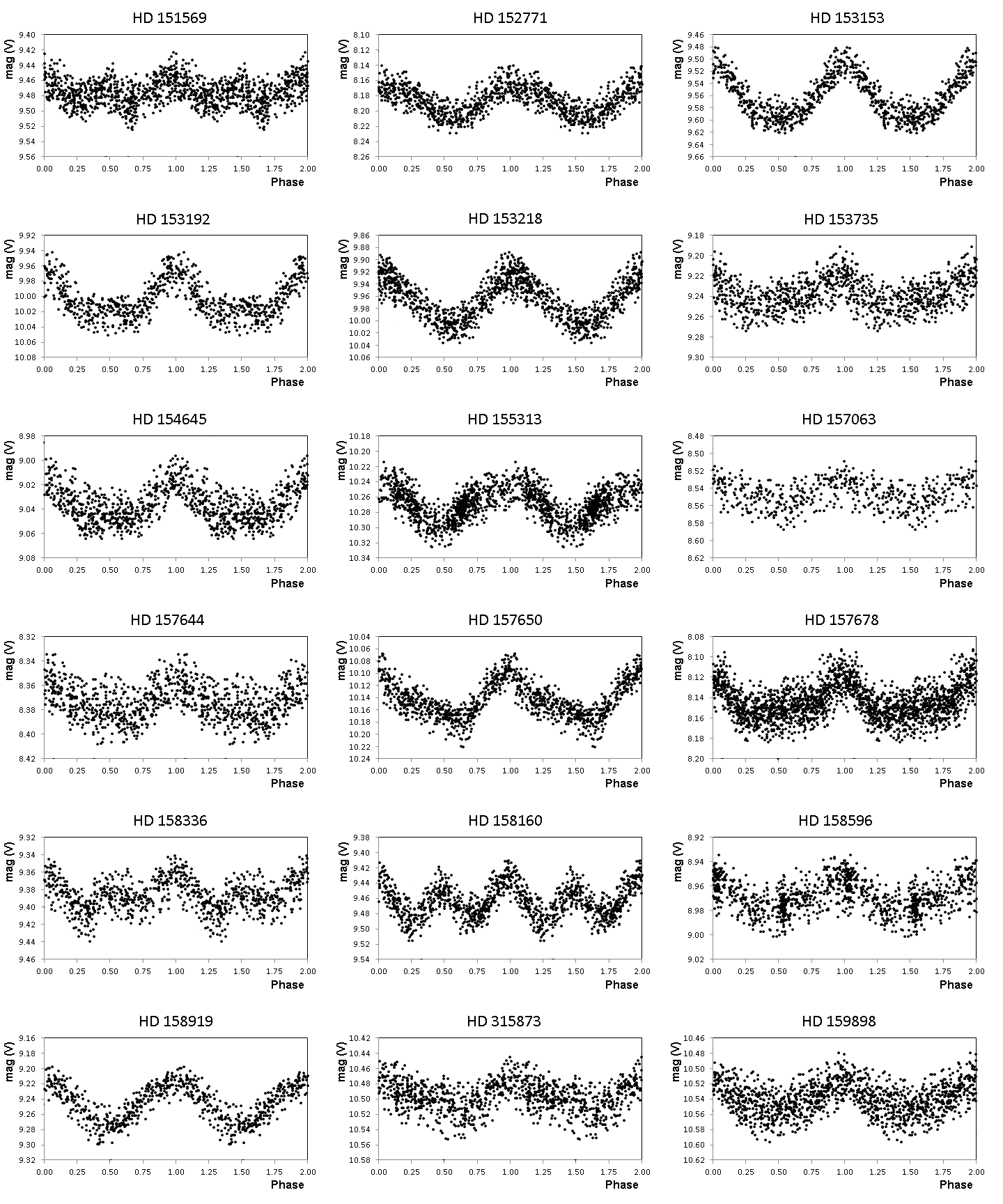}
\caption{continued.} 
\end{center}
\end{figure*}
\setcounter{figure}{0}
\begin{figure*}
\begin{center}
\includegraphics[width=1.0\textwidth,natwidth=1440,natheight=1728]{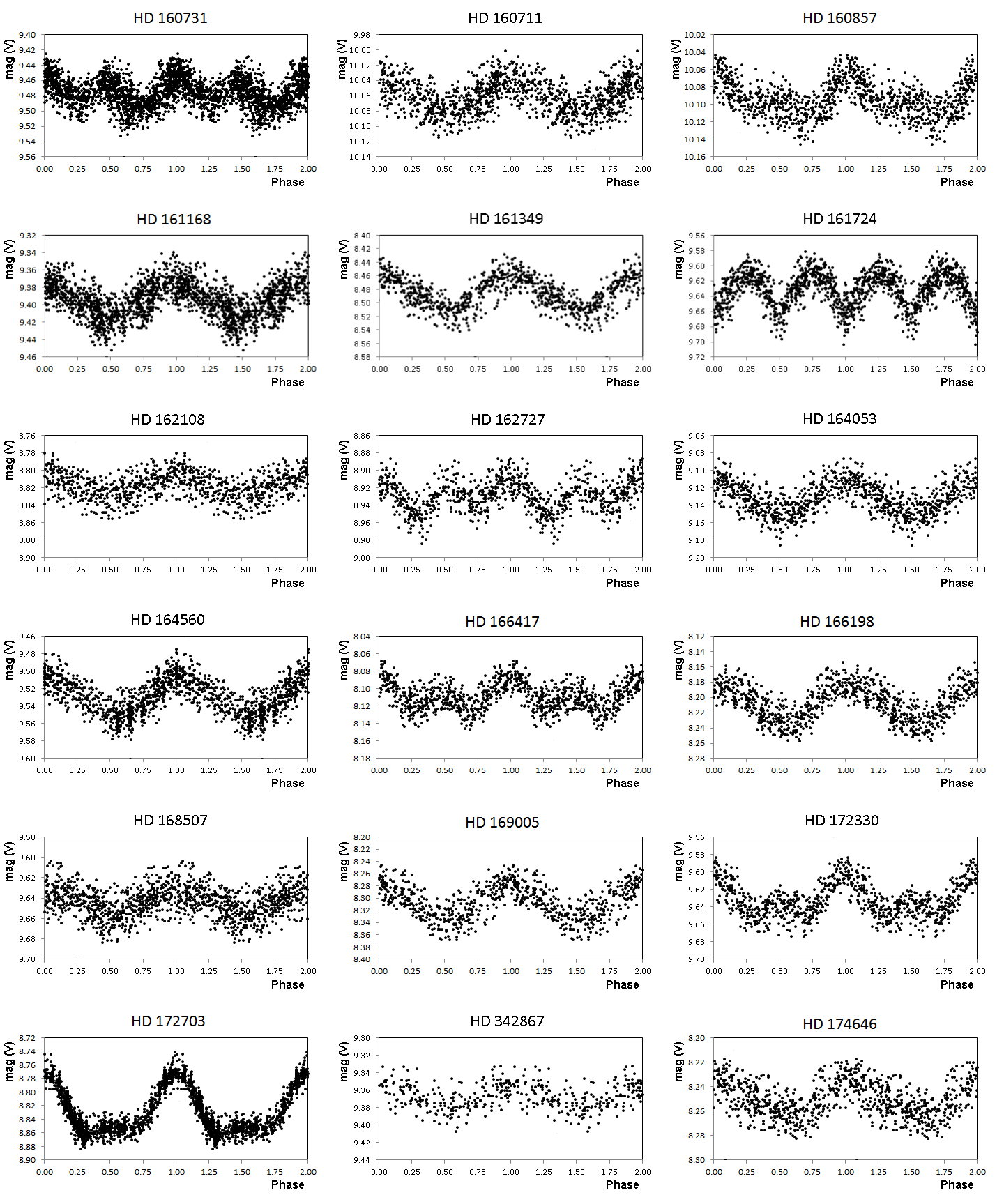}
\caption{continued.} 
\end{center}
\end{figure*}
\setcounter{figure}{0}
\begin{figure*}
\begin{center}
\includegraphics[width=1.0\textwidth,natwidth=1440,natheight=1728]{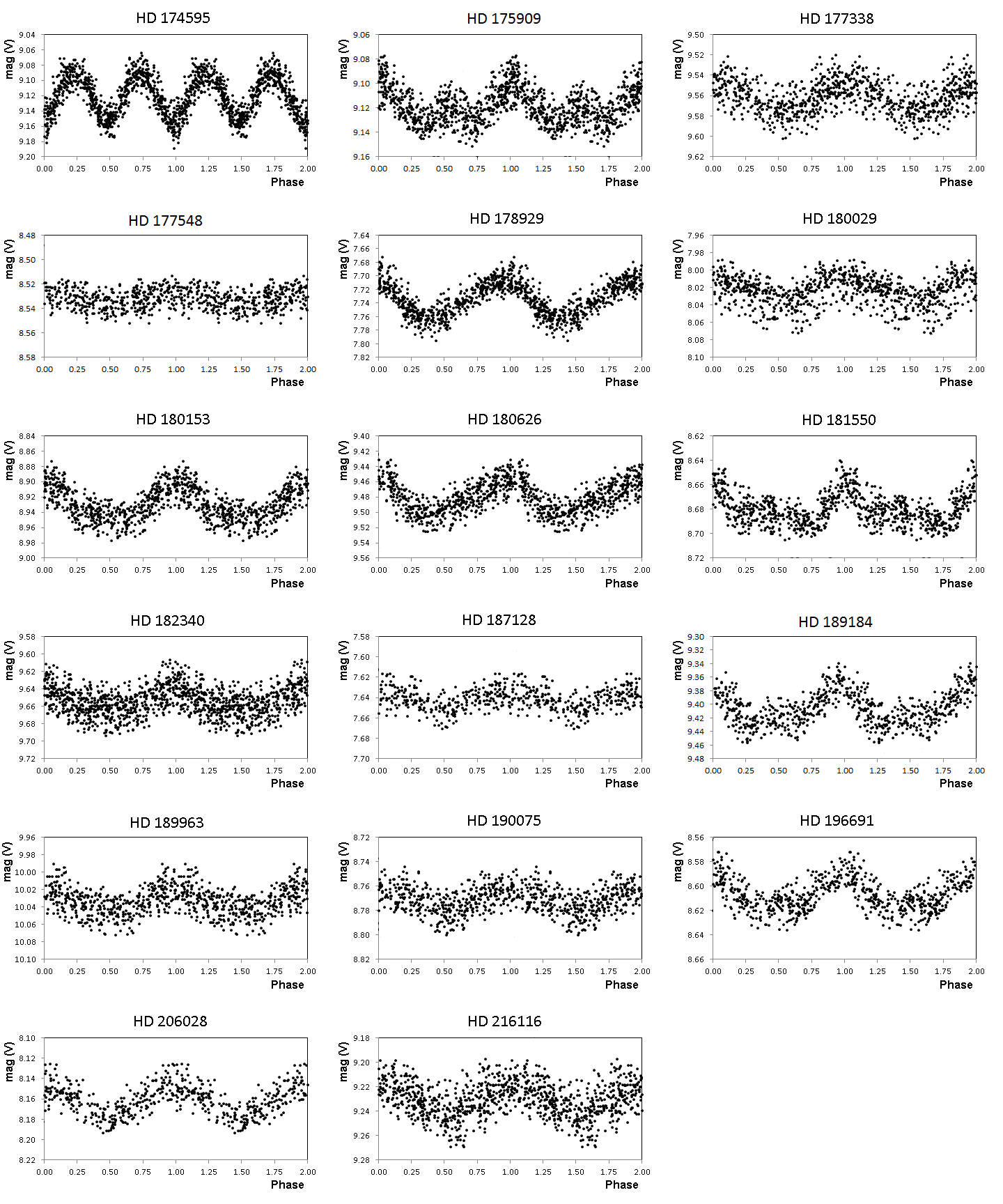}
\caption{continued.} 
\end{center}
\end{figure*}
        
\end{document}